\documentclass[lettersize,journal]{IEEEtran}
\usepackage{amsmath,amsfonts}
\usepackage{algorithmic}
\usepackage{algorithm}
\usepackage{array}
\usepackage[caption=false,font=normalsize,labelfont=sf,textfont=sf]{subfig}
\usepackage{textcomp}
\usepackage{stfloats}
\usepackage{url}
\usepackage{verbatim}
\usepackage{graphicx}
\usepackage{cite}
\usepackage[hidelinks]{hyperref}
\usepackage{flushend}
\usepackage[T1]{fontenc}

\RequirePackage[svgnames]{xcolor}
\usepackage[utf8]{inputenc}
\usepackage{tabu}
\usepackage{booktabs}
\usepackage{gensymb}
\usepackage{siunitx}
\usepackage{pgfplots}
\usepackage{tikz}
\pgfplotsset{compat=newest}
\usepgfplotslibrary{statistics}
\usetikzlibrary{plotmarks}
\usepackage{xcolor}

\pgfmathdeclarefunction{fpumod}{2}{%
    \pgfmathfloatdivide{#1}{#2}%
    \pgfmathfloatint{\pgfmathresult}%
    \pgfmathfloatmultiply{\pgfmathresult}{#2}%
    \pgfmathfloatsubtract{#1}{\pgfmathresult}%
    \pgfmathfloatifapproxequalrel{\pgfmathresult}{#2}{\def\pgfmathresult{5}}{}%
}

\pgfplotsset{ 
    every non boxed x axis/.append style={x axis line style=-},
    every non boxed y axis/.append style={y axis line style=-},
    /pgfplots/boxplot/average={auto},
    boxplot/every average/.style={/tikz/mark=star,}
    }

\begin{document}

\title{An Analysis of Physiological and Psychological Responses in Virtual Reality and Flat Screen Gaming}

\author{Ritik Vatsal*,
Shrivatsa Mishra*,
Rushil Thareja,
Mrinmoy Chakrabarty,
Ojaswa Sharma,
Jainendra Shukla \\Indraprastha Institute of Information Technology, Delhi, India
\thanks{The first two authors (*) contributed equally to this work.}}
        


\maketitle

\begin{abstract}
Recent research has focused on the effectiveness of Virtual Reality (VR) in games as a more immersive method of interaction. However, there is a lack of robust analysis of the physiological effects between VR and flatscreen (FS) gaming. This paper introduces the first systematic comparison and analysis of emotional and physiological responses to commercially available games in VR and FS environments. To elicit these responses, we first selected four games through a pilot study of 6 participants to cover all four quadrants of the valence-arousal space. Using these games, we recorded the physiological activity, including Blood Volume Pulse and Electrodermal Activity, and self-reported emotions of 33 participants in a user study. Our data analysis revealed that VR gaming elicited more pronounced emotions, higher arousal, increased cognitive load and stress, and lower dominance than FS gaming. The Virtual Reality and Flat Screen (VRFS) dataset, containing over 15 hours of multimodal data comparing FS and VR gaming across different games, is also made publicly available for research purposes. Our analysis provides valuable insights for further investigations into the physiological and emotional effects of VR and FS gaming.

\end{abstract}


\section{Introduction}
\IEEEPARstart{E}{motions} in Virtual Reality (VR) tend to be different than those in traditional Flatscreen (FS) gaming because VR provides a more immersive and sensory experience. In traditional FS gaming, players view the game on a two-dimensional screen, which can create a sense of detachment between the player and the game world. However, in VR, players are placed in a fully realized three-dimensional environment and can interact physically with the game world, which creates a stronger sense of presence and immersion \cite{article}. Although prior research has examined the impact of both VR \cite{tan2015exploring, yildirim2018video, rogers2018vanishing, wilson2018violent, granato2020empirical, dey2017effects}, and FS gaming\cite{yannakakis2014emotion, vorderer2004enjoyment,anderson2010violent} on individuals' physiological signals, these investigations have been conducted in distinct settings, leading to different experimental conditions and outcomes. Through our study, the utilization of both measures in conjunction has allowed for the attainment of a comprehensive and holistic perspective on the two distinct gaming methodologies that were subject to testing. This approach has not only allowed us to pinpoint the design obstacles and possibilities that are pivotal to the advancement of the gaming experience but also to obtain a more thorough understanding of the subject matter.

 Recent advances in VR technology have ushered in a new era of VR gaming. As a result, many interactive experiences traditionally available in the conventional  FS space are now moving to VR. Many traditional desktop games, primarily the first-person shooter \cite{tan2015exploring,yildirim2018video} and horror-adventure type games \cite{rogers2018vanishing,wilson2018violent}, have quickly been ported to VR. Therefore, at this time of pivotal change, there is a need to assess the potential usability gains that VR brings to help make quantitative decisions instead of speculative claims. It can be done by investigating the users' emotional responses in VR game-play and by doing a comparative analysis of FS gaming.\par 
 Emotions play a critical part in our lives, influencing our decision-making, perception, social interactions, learning, memory, and creativity  \cite{tripathi2017, zhang2015classification}. Understanding emotional responses helps researchers design better user experiences and improve the usability, acceptability, and accessibility of technologies. We found only a few studies that have explored \textit{emotional challenges} in gaming \cite{Bopp2018, peng2020palette}. Researchers have often used several physiological signals to predict arousal, valence, or even specific emotions by detecting physiological changes in the human body \cite{quesnel_deep_2017}.
 Due to the longer history and greater accessibility, emotions in FS games have been extensively researched, exploring their importance for player engagement and enjoyment. Studies have investigated factors like game content, player experience, personality, and culture, \cite{vorderer2004enjoyment, yannakakis2014emotion, pinitas2023predicting} and the use of physiological and behavioral measures have provided deeper insight into players' emotional responses\cite{anderson2010violent}.

  Our primary novel contribution in this work is to present the first-ever systematic comparison of physiological effects between VR and FS gaming, identifying the emotional and physiological effects of VR and FS gameplay via self-reporting questionnaires and physiological signals. Together, these measures provided a comprehensive subjective and objective outlook on the two gameplay methods tested in our user study. 
  As part of this study, we have curated a 15+ hour multimodal affective dataset, named \textbf{V}irtual \textbf{R}eality and \textbf{F}lat\textbf{S}creen (VRFS). This novel dataset has been made publicly accessible for research purposes. The dataset for this study can be found here - \underline{\href{https://github.com/VREmotions/VRFS}{GitHub link}}. 

\section{Related work} \label{relate_work}
 This study utilizes a methodical approach of literature review that is composed of three separate sections to thoroughly investigate different facets of the research problem. Section ~\ref{emotion_analysis} scrutinizes diverse analytical techniques utilized in emotion research, including physiological measurements and self-report measures. The subsequent section ~\ref{emotions_in_vr} conducts a comprehensive evaluation of the existing literature on emotions in VR. Lastly, section ~\ref{affective_dataset} presents an extensive analysis of recent emotional datasets that are germane to the research and explicates how this study diverges from them, offering a distinctive contribution to the field.

 \subsection{Analysis of Emotions} \label{emotion_analysis}
 Emotional responses can be assessed via three different methods: self-reports from the participant (collected, for example, through questionnaires or interviews), physiological changes in the participant's body (e.g., heart rate or skin conductivity), and directly observable behaviors (e.g., facial expressions or eye gaze) \cite{mauss2009measures}. Among these methods, self-reports are often subject to several biases, and limitations \cite{devaux2016social}, 
  however, these are our greater source for the ground truth. To understand emotions, we must first have a way to quantify them. For this purpose, many researchers have created different methods of classifying emotions, such as the Tree of Emotions \cite{parrott2001emotions}, the wheel of emotions \cite{plutchik2001nature}, the Pleasure, Arousal Dominance scale (PAD), also known as the Valence Arousal Dominance (VAD) model. From the VAD model, the Circumplex Model of Affects (CMA) was derived, which is a two-dimensional Cartesian model to represent emotional stimuli based only on arousal (Y-Axis) and valence (X-Axis) only \cite{russell1980circumplex}. CMA representation is easier to understand and use than the VAD model. The CMA is based on a circular structure, which allows for a more intuitive representation of emotions.\\ 
Additionally, subjective measures like Self-Assessment Manikin (SAM) and Visual Analog Scale (VAS) are commonly employed to assess arousal and valence in VR \cite{10.1145/3383812.3383844, GANRY2018257}. SAM and VAS offer a convenient and reliable way to quantify emotional responses, aiding researchers in understanding the emotional impact of VR environments and applications.
  
 Meanwhile,  emotion recognition from physiological signals is expedient since it taps the pure, unaltered emotion in contrast to behavioral responses like facial expressions, which can be faked. Recent advancements in wearable technologies have shown a strong potential for hassle-free acquisition of physiological signals in a non-intrusive manner and have thus inspired us to investigate emotional responses in VR and FS gameplay using physiological signals. Physiological signals offer additional unique information regarding users' interactions with and responses to a system. Researchers have also utilized multiple sensors to help evaluate and monitor human physiological signals. Signals like EDA \cite{macedonio2007, valenza2012}, ECG \cite{nardelli2015, panda2020}, and Electroencephalogram (EEG) \cite{jalilifard2016, kosunen2016, tripathi2017} have been widely used. The use of these sensors is showcased in a wide variety of applications, including medical \cite{plawiak2018novel, greco2016advances}, neuromarketing \cite{yadava2017analysis, cuesta2018neuromarketing}, sports training \cite{sessa2018sports, butkevivciute2019removal} and many more.
Additionally, researchers have also proposed the use of these physiological signals as controllers in games to enhance connect among multiple players \cite{10.1145/3290607.3313268}. Researchers have used physiological signals like electrodermal and cardiovascular activity to analyze and predict emotions \cite{7394401}. They have used these findings to create guidelines for VR developers to create more immersive games. Classification and analysis of emotions have been important for researchers for a long time. From a tree-like structure that divided emotions into primary, secondary, and tertiary \cite{parrott2001emotions} to a size axis model proposed by Kort et al. \cite{kort2001affective}. This led to the development of the Valence Arousal and Dominance Model by Osgood, Suci, and Tannenbaum \cite{osgood1957measurement}. 

 The use of physiological signals for emotional classification has increased over the past decade thanks to improved technology surrounding procuring such data. For instance, The recent Empatica E4 device has shown the ability to capture physiological data reliably in most cases \cite{mccarthy2016validation}.
 \subsection{Emotions in VR} \label{emotions_in_vr}
 Emotional analysis for immersion in VR has been much studied recently. Driving scenarios are commonly used in VR research as a means to study and understand human perception, due to the familiar and relatively realistic experience they offer. Researchers have used driving simulation games and wheel controllers to create an environment as close as possible to real driving and used self-reporting surveys for felt emotions, and experience \cite{10.1145/3027063.3053202, Cao2020, dey2017effects}. These studies revealed that VR dissociates the user from the real world more than FS, thus making VR more immersive. The uneasiness and motion sickness caused by the current VR systems made these studies more challenging. Research with young adults on a driving simulator in VR and FS showed that VR elicits more positive emotions, and the sense of immersion and flow is more remarkable in VR games than in FS \cite{pallavicini2019comparing}. Researchers have established that EEG signals from the brain are a relevant metric for decoding emotional arousal \cite{hofmann2021decoding}. One study tried to recreate traffic-light-based responses in VR and tracked the system's effectiveness with EEG signals from the brain \cite{lin2007eeg}. They showed that in an average of eight participants, traffic events could reach an 87\% accuracy compared to real-world events in VR. Studies have also shown that VR can elicit specific emotions with predictable outcomes. Researchers have been able to create "anxious" and "relaxation" producing simulations in VR with a significant certainty of predicted emotion arousal \cite{riva2007affective}. Further, they have compared these simulations between VR and FS, and have found VR to be more "relaxing" and immersive than FS \cite{reece2022exposure, knaust2022exposure}. With these results, researchers have looked into VR as a medium for required emotion elicitation as an upgrade to existing stimuli. Meuleman et al. \cite{meuleman_induction_2021} observe that emotional responses in a VR gaming study are clustered in two segments; joy and fear. Work has also been done to successfully create deep learning models to predict peak emotions in gameplay using several biosignals \cite{quesnel_deep_2017}. VREED \cite{tabbaa_vreed_2022}, is one of the first datasets that provide behavioral (eye tracking) and physiological signals [Electrocardiogram: ECG and Galvanic Skin Response: GSR (also known as Electrodermal Activity: EDA)] data in addition to self-reporting for comparison of emotion elicitation in VR and FS 360° Video-Based Virtual Environments.
Further, recognizing the emotions users feel allows researchers to control these scenarios live, making systems like this useful for emotional training for children with Autism. Similar research has been done to compare the immersiveness and motion sickness in first and third-person points of view in VR \cite{https://doi.org/10.1002/cav.1830}, which suggests that although the first-person perspective is more immersive, it is also more prone to inducing sickness. The emotions induced and their strength varies significantly in FS and VR as factors such as screen size and worldview have been shown to significantly affect mental immersion, which in turn affects emotions felt by the users \cite{schmidt2020,shelstad2017gaming,10.1145/3411764.3445492}. Other factors such as \textit{cybersickness} also come into play when exploring emotions in VR gaming \cite{YILDIRIM201935}. Overall positive emotions such as \textit{hope, courage, relaxation, and calmness} are more associated with gaming in VR \cite{peng2020palette}; however, it is essential to note that these emotions are also greatly affected by the genre of the game title. It is possible to induce particular emotions by designing gameplay in a particular manner \cite{friedrichs2015simple}.

 \subsection{Affective Datasets} \label{affective_dataset}
 Data gathering is essential for many studies; however, this process is extremely lengthy, expensive, and requires significant resources. To combat these, many studies utilize publicly available datasets instead. One prominent dataset is the International Affective Picture System (IAPS) \cite{lang2005international} that provides a range of picture-based emotional stimuli and participant ratings. Another recent example is the expanded version of the International Affective Digitized Sounds (IADS-E) \cite{yang2018affective} dataset. It augments the previous International Affective Digitized Sounds (IADS) \cite{bradley2007international} dataset and provides a larger selection of audio-based emotional stimuli combined with the participant rating.
 
 Instead of just the emotional ratings, many datasets also offer physiological as well as behavioral data of the participants \cite{10.1145/3410404.3414227, 6553805, 8854185}. These datasets record the participants' behavioral (eye gaze movement, face recording) and physiological (EDA, Blood Volume Pulse: BVP, voice features) signals while they are exposed to emotional stimuli. For example, MAHNOB-HCI \cite{manhob-hci} is an affective dataset consisting of the face videos, audio signals, eye-gaze data, and peripheral/central nervous system physiological signals of participants reacting to 20 different emotional videos providing a valuable resource for analyzing emotion recognition and affective computing. Another dataset called the Database for Emotion Analysis using Physiological signals (DEAP) \cite{DEAP} analyzes affective music videos, their reported ratings, and the recorded EEG data of the participants. Similarly, the Database for Emotion Recognition through EEG and ECG Signals from Wireless Low-cost Off-the-Shelf Devices (DREAMER) \cite{DREAMERS} analyzes the ECG and EEG data from participants experiencing audio-visual stimuli. The MEG-Based Multimodal Database for Decoding Affective Physiological Responses (DECAF) \cite{DECAF} contains the recorded brain signals similar to DEAP. However, these were captured using a Magnetoencephalogram (MEG) that requires physical contact with the participant's scalp but facilitates a natural affective response.
 
 There also exist multiple datasets for participants reacting to different stimuli, such as the Bio-Reactions and Faces for Emotion-based Personalization for AI Systems (BIRAFFE2) \cite{kutt2022biraffe2} that consists of accelerometer data, ECG, and EDA signals, participants’ facial expression data, and a personality and game engagement questionnaires. In the field of VR, VREED \cite{tabbaa_vreed_2022} is one of the first datasets that analyzes behavioral (eye tracking) and physiological signals (ECG and EDA) data in addition to self-reporting for comparison of emotion elicitation in VR and FS $360\degree$ in video-based virtual environments. Another dataset, Affective Virtual Reality System (AVRS) \cite{zhang2017affective}, has rated arousal, valence, and dominance of similar 360° VR environments using self-reporting. Further, datasets have rated arousal and valence and have analyzed correlations between head movement and self-reported values \cite{li2017public, singh2023have}. In addition to self-reporting, some datasets have also used physiological data collected via EEG for emotional classification \cite{bdcc6010016, horvat2018assessing}.

\bigskip

The review outlined in section ~\ref{emotions_in_vr} establishes that there is a dearth of rigorous analysis regarding the physiological effects associated with VR and FS gaming. This deficiency in the study of physiological and psychological reactions to both VR and FS gaming can be attributed to the unavailability of standardized datasets that offer emotional responses for both types of gaming. Section ~\ref{affective_dataset} of the review highlights the significant limitation of many affective datasets, including DEAP \cite{DEAP} and MAHNOB-HCI \cite{manhob-hci}, in that they do not utilize VR stimuli to elicit emotions, despite its well-known ability to provide immersive experiences. Additionally, existing datasets that analyze VR stimuli rely on different stimulus modalities, such as 360° videos [38] or pictures embedded in VR environments such as $360\degree$ videos \cite{li2017public} or pictures embedded in VR environments \cite{horvat2018assessing}. Most affective datasets primarily focus on either physiological or psychological responses, making systematic comparisons between VR and non-VR responses challenging and hampering the ability to draw comprehensive conclusions. This limitation emphasizes a gap in the field and presents an important opportunity for researchers to perform more comprehensive and standardized investigations that could improve the understanding of individuals playing games in VR and aid in enhancing the experience for new users. To this end, we have also made the entire VRFS dataset publicly accessible.

\section{Stimuli Selection Procedure}\label{Stimuli_Selection_Procedure}
The selection of effective games is essential to elicit the appropriate affective responses; therefore, our selection process underwent multiple stages to select the best games. We first shortlisted commercially available games that had both VR and FS variants available, based on online reviews, and placed them on a CMA. After placing games on the CMA, 12 games were shortlisted such that each quadrant of the CMA had three games. This pilot study selects one game from each quadrant of the CMA. Our game selection process is summarized below.
\begin{itemize}
\item [] \textbf{Focus groups}
    \begin{itemize}
        \item Three Researchers met over two 2-hour sessions to extensively discuss and experience the 12 games to find suitable games for the pilot trial.
    \end{itemize}
\item[] \textbf{Pilot trial}
    \begin{itemize}
        \item In a span of two weeks, six volunteers spent around 90 minutes each engaging in and rating the selected 12 games. 
        \item Each volunteer played four games, such that each game was played twice.  
        \item SAM [arousal, valence, dominance] and VAS [joy, anger, calmness, sadness, disgust, relaxation, happiness, anxiousness, fear, and dizziness] were recorded to select the game.
    \end{itemize}
\end{itemize}
\vspace{-6pt}
\subsection{Stimuli Shortlisting}
Using various game distribution platforms such as Steam and Microsoft Xbox, the researchers attempted to find games that offered the same gameplay in both FS and VR modes. A short list was curated according to the following inclusion criteria:
\begin{itemize}
  \item Games that were available on the Microsoft Windows platform,
  \item Games that offered similar experiences both in FS and VR, apart from obvious medium differences such as resolution and controls,
  \item Games that had uniform and predictable play sessions (e.g. Racing games with fixed laps or simple task-based games) for all participants,
  \item Games from which a sub-section can be identified that can be replayed or recreated by participants, and
  \item Games that could run comfortably on the test system (consistent frame rate with no frame drops).
\end{itemize}
After shortlisting, the researchers found 21 games and selected 12 based on each game's position in the CMA. The aim was to have three games in each quadrant. These 12 games were used for the pilot trial and then narrowed down to 4, one from each quadrant.

\subsection{Pilot Trial}
Six volunteers (five males and one female) aged between 19 and 21 ($\mu=20$, $\sigma=0.74$) played and rated the selected 12 games. The volunteers played four games each in a  randomized order. Each game was effectively played and rated twice. To optimize the resources, each volunteer only played the FS version of each game. Since the perceived category of emotions has not been found to differ significantly between VR and FS (for example, see \cite{al2021comparison}), we expected that the chosen games through such a pilot trial would elicit the desired emotions in VR gameplay as well. The following tools were used to judge arousal:
\begin{itemize}
    \item  SAM is a widely used state measurement tool \cite{bradley1994measuring}. It has simple cartoon-like manikin icons that can be used to plot the CMA dimensions (arousal and valence). The valence scale ranges from 1=“happy” to 5=“sad” pictures of SAM. Meanwhile, the arousal scale ranges from 1=“calm” to 5=“excited” pictures of SAM.
    \item VAS \cite{hawker2011measures} is a continuous slider-based scale of numbers ranging from 1-100, with two verbal descriptors at each end. Using VAS, volunteers rated how they felt while engaging in the games,
    using separate scales for joy, happiness, calmness, relaxation, anger, disgust, anxiousness, fear, and sadness.
\end{itemize}

After completing the pilot study, we had the SAM and VAS results for each game, using which we could see how each game performed and in which quadrant they lie, as shown in Fig.~\ref{fig:pilot_quad}. Arousal and valence ratings for each of the 12 games can be found in Table~\ref{tab:PilotTable}.  SAM and VAS ratings have been collected to be used as the "ground truth" or reference point for the research, against which the physiological responses can be compared. Although we collected dominance as well through SAM, we found that it did not differ significantly between participants (H: 11.0, p-value: 0.44) according to the Kruskal Wallis test, so we excluded the dominance while performing the stimuli selection.

\begin{table}[h!]
\caption{Reported Arousal and Valence of Each Game in the Pilot Study. The Selected Games are Highlighted in Bold.}
\label{tab:PilotTable}
\begin{center}
\begin{tabular}{lccS[table-format=1.2]S[table-format=2.1]} \toprule
    \textbf{Name} & \textbf{Arousal} & \textbf{Valence}& \textbf{Distance $r$} & \textbf{$d\theta$} \\\midrule
    \textbf{War Thunder} & 2 & 3 & 0.7  & 2.8\\
    Hitman & 3 & 4 & 1.58  & 27.8\\
    Raceroom & 3.5 & 3 & 1.11  & 14.4\\
    \textbf{Minecraft} & 1 & 2 & 1.58  & 22.0 \\
    No Man's Sky & 1.5 & 2.5 & 1.0  & 45.0\\
    The Forest & 3.5 & 2.5 & 1.0  & 45.0\\
    Forewarned & 2.5 & 1.5 & 1.0  & 45.0\\
    Project Cars 2.0 & 2 & 3.5  & 1.11 & 20.1\\
    Microsoft Flight Sim & 2 & 2.5 & 1.0 & 45.0\\
    Subnautica & 3 & 1  & 2.06  & 31.8\\
    \textbf{Phasmophobia} & 3.5 & 2  & 1.11  & 14.5 \\
    \textbf{Dirt Rally 2.0} & 4 & 3.5  & 1.80 & 7.7\\\bottomrule
\end{tabular}
\end{center}
\end{table}

\newcommand*{\XAxisMin}{0}
\newcommand*{\XAxisMax}{5}
\newcommand*{\YAxisMin}{0}
\newcommand*{\YAxisMax}{5}
\begin{figure}
    \centering
        \begin{tikzpicture}[thick,scale=0.7, every node/.style={scale=0.7}]
        \pgfplotsset{set layers}
        \begin{axis}[
            axis equal image, clip=false,            
            legend style={legend columns=1, 
            legend style={at={(1.5,0.507)},anchor=south west}}, legend cell align={left},
            xmin=\XAxisMin, xmax=\XAxisMax, ymin=\YAxisMin, ymax=\YAxisMax,
            width=0.6\textwidth
        ]
        \addplot+[only marks, mark=diamond*] coordinates{(3, 2)}; \addlegendentry{War Thunder}
        \draw[red, thick] (axis cs:3, 2) circle [radius=0.2];          
        \addplot+[only marks, mark=pentagon*] coordinates{(4, 3)}; \addlegendentry{Hitman}
        \addplot+[only marks, mark=Mercedes star] coordinates{(3, 3.5)}; \addlegendentry{Raceroom}
        \addplot+[only marks, mark=triangle*] coordinates{(2, 1)}; \addlegendentry{Minecraft}
        \draw[red, thick] (axis cs:2, 1) circle [radius=0.2];          
        \addplot+[only marks, mark=heart] coordinates{(2.5, 1.5)}; \addlegendentry{No Man's Sky}
        \addplot+[only marks, mark=otimes*] coordinates{(2.5, 3.5)}; \addlegendentry{The Forest}
        \addplot+[only marks, mark=oplus*] coordinates{(1.5, 2.5)}; \addlegendentry{Forewarned}
        \addplot+[only marks, mark=star] coordinates{(3.5, 2)}; \addlegendentry{Project Cars 2.0}
        \addplot+[only marks, mark=o] coordinates{(2.5, 2)}; \addlegendentry{MS Flight Sim}
        \addplot+[only marks, mark=10-pointed star] coordinates{(1, 3)}; \addlegendentry{Subnautica}
        \addplot+[only marks, mark=square*] coordinates{(2, 3.5)}; \addlegendentry{Phasmophobia}
        \draw[red, thick] (axis cs:2, 3.5) circle [radius=0.2];        
        \addplot+[only marks, mark=] coordinates{(3.5, 4)}; \addlegendentry{Dirt Rally 2.0}
        \draw[red, thick] (axis cs:3.5,4) circle [radius=0.2];
        \addplot[mark=none, black] coordinates {(\XAxisMin, 2.5) (\XAxisMax, 2.5)};
        \addplot[mark=none, black] coordinates {(2.5, \YAxisMin) (2.5, \YAxisMax)};
        \node[left = 10pt] at (0, 3.75) {\rotatebox[origin=c]{90}{High Arousal}};
        \node[left = 10pt] at (0, 1.25) {\rotatebox[origin=c]{90}{Low Arousal}};
        \node[below = 10pt] at (3.75, 0) {High Valence};
        \node[below = 10pt] at (1.25, 0) {Low Valence};
        \end{axis}
        \end{tikzpicture}
    \caption{Games plotted in respective quadrants according to arousal and valence results from pilot SAM study. Encircled games were selected for the final study.}
    \label{fig:pilot_quad}
\end{figure}
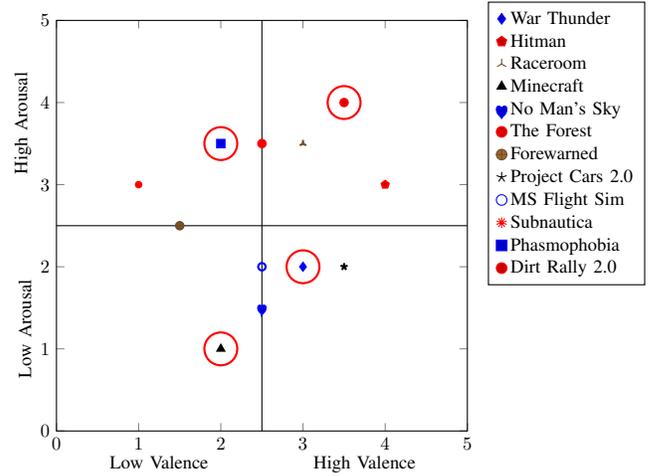

\begin{figure*}[h!]
    \centering
    \begin{tabular}{cc}
    \includegraphics[width=0.4\textwidth, height=0.2\textwidth]{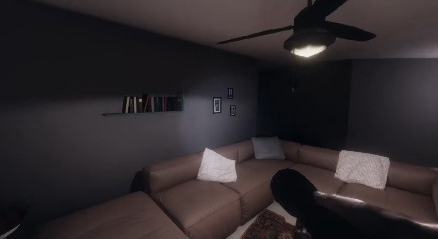} &
    \includegraphics[width=0.4\textwidth, height=0.2\textwidth]{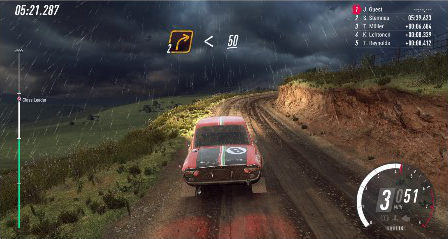} \\
    (a) & (b)\\
    \includegraphics[width=0.4\textwidth, height=0.2\textwidth]{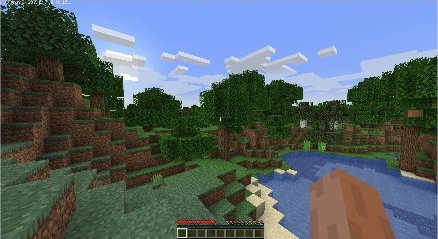} &
    \includegraphics[width=0.4\textwidth, height=0.2\textwidth]{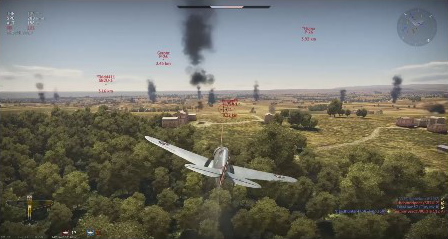} \\
    (c) & (d)
    \end{tabular}
    \caption{In-game screenshots of selected four games. (a) Phasmophobia: High Arousal Low Valence; (b) Dirt Rally 2.0: High Arousal High Valence; (c) Minecraft: Low Arousal Low Valence; (d) War Thunder: Low Arousal High Valence}
    \label{fig:In-Game Screen Shots}
\end{figure*}

In order to  analyze  which game should be chosen as the representative of their quadrant, we plotted (see Fig.~\ref{fig:pilot_quad}) the distance $r$ of that game from the origin as well as the angle $d\theta$ that they are farther from $45\degree$ as
\begin{equation}
    r = \sqrt{(|2.5-Arousal|)^2 + (|2.5-Valence|)^2}, \text{ and}\nonumber
    \label{eq:r}
\end{equation}
\begin{equation}
    d\theta = \bigg |\frac{\pi}{4} - \tan^{-1}{\left(\frac{(|2.5-Arousal|)}{(|2.5-Valence|)}\right)} \bigg | \nonumber
    \label{eq:dt}
\end{equation}
Games with the highest $r$ value, i.e., the largest distance from the origin and the least difference from $45\degree$, would be the best representatives for that quadrant. Taking values farther away from the origin would show higher arousal or valence in that preferred quadrant. Values near the $45\degree$ line would ensure that both arousal and valence are almost equally high. The following games were chosen from each quadrant (see gameplay screenshots in Fig.~\ref{fig:In-Game Screen Shots}). 
\begin{itemize}
\item \emph{Quadrant 1: High Arousal High Valence (HVHA)} Dirt Rally 2.0 (Fig.~\ref{fig:In-Game Screen Shots}(b)) is chosen as it has the minimum $d\theta$ value of 7.7 along with the maximum $r$ value of 1.80. This is a rally-style racing game that offers a realistic driving experience.
\item \emph{Quadrant 2: High Arousal Low Valence (LVHA)} Phasmophobia (Fig.~\ref{fig:In-Game Screen Shots}(a)) is chosen in this quadrant due to its high $r$ value of 1.11 and is the closest to $45\degree$ with a $d\theta$ value of 15.5. This is a First-Person horror game where the users have to find clues to solve an objective.
\item \emph{Quadrant 3: Low Arousal Low Valence (LVLA)} Minecraft (Fig.~\ref{fig:In-Game Screen Shots}(c)) is the clear winner here since it has a higher $r$ value of 1.58 with a minimum $d\theta$ value of 22. This is a First-Person block-based sandbox video game in which players gather and use blocks to create various objects.
\item \emph{Quadrant 4: Low Arousal High Valence (HVLA)} We chose War Thunder (Fig.~\ref{fig:In-Game Screen Shots}(d)) in this quadrant since it has a much lower $d\theta$ value of 2.8 compared to other games. This military-style aerial combat fighting simulator aims to shoot down enemy aircraft.
\end{itemize}

\section{Experimental Setup and Data Collection}\label{Experimentall_Setup}
 This experimental study aimed to gather physiological and self-reporting data from participants in different emotion-eliciting games played by them in both VR and FS mediums. Further, we want to analyze and identify these signals for evident patterns of arousal and immersiveness and report our findings to future researchers and developers.

\subsection{Ethics}
 The study was approved by the Institutional Review Board (IRB) of Indraprastha Institute of Information Technology, Delhi. All participants were provided with an overview of the study and a written consent was taken before the study was conducted. To preserve the privacy of all participants, raw physiological data collected during the study would not be available. Only pre-processed and anonymized data would be shared publicly.

\subsection{Participant Screening Criteria}
 An invite was sent via various channels to students of our institute. The invite had a brief description of the experiment and a demographic questionnaire. The following volunteers were excluded from the study: 
\begin{itemize}
    \item Individuals who reported having a seizure(s), migraines, or any other medical condition in the past that
could increase the risk during the user study, 
    \item Participants who reported having a prior history of motion sickness,
    \item Individuals who rated four or higher on a Likert scale on the question ``How easily do you get motion sick or carsick?'' where one was ``Rarely get motion sick'' and five was ``Very often get motion sick'', and
    \item Due to the ongoing COVID-19 pandemic, participants who were COVID positive or reported symptoms like cough, fever, and fatigue.
\end{itemize}
 Finally, 36 individuals (16 female and 20 male) aged between 19 and 22 years ($\mu = 20.8, \sigma=0.71$) volunteered to participate in this study. 45.4\% of the participants (n=15) reported having used VR previously (Responded 'Yes' to 'Have you experienced VR before?'), and none reported feeling any motion sickness during or after exposure to VR.

\subsection{Psychological Measures}
Participants completed the following self-reported measures during the experiment:
\begin{itemize}
    \item \emph{A pre-exposure questionnaire} 
    \begin{itemize}
        \item Participants were asked to report any demographic information such as age, gender, ethnicity, dominant hand, and previous experience with VR.
    \end{itemize}
    \item \emph{A post-exposure questionnaire}
    \begin{itemize}
        \item SAM and VAS measures were taken immediately after playing the game in both VR and FS
        \item Using VAS, participants were asked to rate how dizzy they felt after playing the games in both VR and FS.
    \end{itemize}
\end{itemize}

\subsection{Physiological Measures}
 With the four selected games from the pilot study, we invited 33 participants to play these games in FS and VR. Each participant only played one game to prevent VR fatigue. The details of the data we collected are explained in Table~\ref{tab:exp}. We measured the participants' BVP and EDA. These signals have been shown to represent physiological state accurately \cite{tabbaa_vreed_2022} and can be measured non-intrusively with a wrist device. The device is worn similarly to a wristwatch and does not obstruct the gameplay experience of the participants. The BVP and EDA signals are explained in more detail below. These signals were recorded using an Empatica E4 device and synced with each other. The sampling rate for these is hard coded into the firmware and optimized to capture the frequency content of relevant signals (BVP - 64 Hz, EDA - 4 Hz).
\subsubsection{BVP}
 The BVP signal measures the changes in blood volume inside arteries and capillaries. A Photoplethysmography (PPG) sensor uses light from an LED to measure the refraction and estimate blood volume. The amount of light that returns to the PPG sensor is positively proportional to the blood volume. PPG gives the average value of blood in the tissues through which light has passed. The BVP in Empatica E4 is processed through the  PPG  sensor at 64 Hz. Using this measure of the amount of blood passing over each pulse through a PPG, we can effectively calculate Heart Rate variability (HRV) using BVP \cite{2639332720070101}. We then extract the Low-frequency (LF) and High-frequency (HF) features from the HRV. The LF band (0.04–0.15 Hz) is affected by breathing from ~3 to 9 bpm. \cite{kamath2012heart} while the HF or respiratory band (0.15 – 0.40 Hz) is influenced by breathing from 9 to 24 bpm \cite{task1996heart}. The ratio of LF to HF power (LF/HF ratio) has been proposed as an indicator of sympathovagal balance, reflecting an increase in values during states of “sympathetic dominance” induced by emotional and physiological stress \cite{eckberg1997sympathovagal}. Similar to previous research\cite{shaffer2014healthy}, we have employed the LF/HF ratio within the context of an ambulatory condition, which encompassed participants having unrestricted movement (with the exception of keeping the off-hand restricted) and accounting for the variability in breathing patterns.

\subsubsection{Electrodermal Activity}
 Electrodermal activity (EDA) refers to the variation of the electrical properties of the skin in response to sweat secretion. Applying a low constant voltage,  Skin Conductance  (SC) change can be measured non-invasively. EDA has been measured with the Galvanic Skin Response sensor. EDA was sampled at a frequency of 4 Hz, with the data being measured in micro siemens ($\mu$S). EDA data consists of two components, Skin Conductance Level (SCL) [tonic component] and Skin Conductance Response (SCR) [phasic component]. SCL is a general measure of slow-moving psycho-physiological activation \cite{setz2009discriminating}, while SCRs depict higher-frequency changes directly related to an external stimulus \cite{greco2015cvxeda}. The time series of SC can be characterized by a slowly varying tonic activity (i.e., SCL) and a fast varying phasic activity (i.e., SCR) \cite{BENEDEK201080}.

\begin{table}[!h]
\begin{center}
\caption{ Details of Experimental Data Collection. }
\label{tab:exp}
\begin{tabular}{p{0.12\textwidth} p{0.32\textwidth}} \toprule
 Stimuli Selection Method 
 & Based on the six pilot trial results\\
 \midrule
 Participants Information and Pre-Exposure Data & 33 Participants provided demographic information (i.e. sex, age), answered a health inclusion questionnaire\\
 \midrule
 Post-Exposure Data & SAM (arousal, valence) and VAS\\
 \midrule
 Recorded Physiological Signals & BVP and EDA\\
 \midrule
  Study Design &  2x4 mixed-participant study (2 conditions (VR+FS) $\times$ 4 variables (4 Games))\\
  \midrule
 Additional Materials & Self Reporting Questionnaire, and verbal instructions protocol\\\bottomrule
\end{tabular}
\end{center}
\end{table}
\vspace{-15pt}
\subsection{Apparatus and Setup}
All apparatus was set up in the ``Experiment Room,'' and only one participant and two researchers were allowed at a time to ensure necessary social distancing rules. Following are various  hardware devices  that were part of our experimental setup:
\begin{itemize}
\item \textit{ FS Display and Speakers:} A single HP 24-inch Monitor was used to play games on the FS. The monitor resolution was 1920x1080 pixels. The refresh rate was 60 Hz, and all FS games were played at 60 FPS to ensure a smooth experience. The monitor comes with built-in speakers, which were used to stream audio in FS mode.
\item \textit{ VR Headset:} An Oculus Rift S VR headset\footnote{Oculus Rift S VR headset: https://www.oculus.com/rift-s/} was used for the VR experience. It comes with one headset and two controllers. The headset is wired to the computer, and both handheld controllers are wireless. Rift S uses a single fast-switch LCD panel with a resolution of 2560×1440 pixels and an 80 Hz refresh rate. All VR games were played at 80 FPS. It has a field of view of 115 degrees. The Rift S has inbuilt speakers in the headset, positioned just above the ears, which are used for audio delivery in VR experiences.
\setlength\parindent{12pt}\textit{ Steering Wheel Controller: } For the driving game (Dirt Rally 2.0), we used a steering wheel controller Thrustmaster T300RS\footnote{Thrustmaster T300RS Force Feedback racing wheel: https://www.thrustmaster.com/en-gb/products/t300rs/}. This is a steering wheel and a 2-leg paddle device. This simulates a real-world driving experience. The wheel also has vibrational feedback.
\item \textit{ Physiological Collection Device:} The Empatica E4\footnote{Empatica E4 Wristband: https://www.empatica.com/research/e4/} wristband was used to measure various physiological signals. It has sensors for PPG (measures  BVP), a 3-axis accelerometer, an EDA sensor, and an infrared thermopile for skin temperature. It also has an internal clock for syncing and a physical event marking button to mark events in real time.  
\end{itemize}

Both VR \& FS games were played on a system above recommended specifications: Intel i7 8700K CPU, NVidia GTX 1080Ti GPU. In addition, a 14-inch FHD Windows laptop was used for self-reported emotion questionnaires (SAM \& VAS).

\subsection{Hypotheses}
Overall, we expected VR games to be perceived as more immersive \cite{pausch1997quantifying} and arousing. Due to the high physical nature of control in VR games, we expected the mentally and physically challenging nature to reflect in the physiological signals. The following hypotheses informed our study. 
\begin{enumerate}
    \item\textbf{H1:} Previous research has suggested a positive correlation between heart rate and cognitive load \cite{held2016bimodal, can2019stress, malinska2015heart}. Moreover, studies have shown higher immersion and increased cognitive load in virtual reality (VR) gaming compared to traditional first-person shooter (FS) gaming \cite{blascovich2011infinite}. Consequently, we anticipated an elevated heart rate during VR gaming.
    
    \item\textbf{H2:} A positive relationship is observed between emotional intensity and immersion \cite{riva2007affective}.  Since immersion in VR is higher than in FS gaming, we expected that immersion induced through VR gameplay would enhance one’s appraisal of their context compared to FS gameplay.
    
    \item\textbf{H3:} Due to the higher cognitive load involved, VR gaming induces greater emotional and physiological stress compared to FS gaming\cite{conway2013effect}. Additionally, emotional and physiological stress often leads to an increase in LF/HF sympathovagal balance values, indicating sympathetic dominance \cite{eckberg1997sympathovagal}. Hence, we anticipated an elevation in LF/HF ratio during VR gaming.

    \item\textbf{H4:} A higher cognitive load is observed with learning novel information \cite{sweller2011cognitive}. VR controls and navigation are relatively new to the target users, requiring more concentration and physical effort. Accordingly, users are likely to feel less dominant during VR gaming compared to FS gaming. 
    
    \item\textbf{H5:} VR gaming is likely to induce more arousal compared to FS gaming due to higher immersion in the VR environment, which will be reflected in increased SAM ratings. Since SCR and SCL have been found to be much more frequent when the individual is aroused \cite{boucsein2012electrodermal}, we expected an increased skin conductance (EDA) activity as well.

    \item\textbf{H6:} Past literature establishes a correlation among valence, arousal, and dominance\cite{warriner2013norms}. In the present context, we postulate a heightened prominence of this correlation within Virtual Reality (VR) gaming as compared to Full-Screen (FS) gaming.
    
\end{enumerate}

\subsection{Experimental Procedure}
 We collect and analyze qualitative psychological responses and quantitative physiological measures in a 2x4 mixed-participant study. Each session lasted between 30-40 minutes, depending on the game. Each participant was asked to visit the lab twice and play the same game on the FS in one session and VR in the other, with a gap of 45-60 minutes between sessions. To avoid the order effects, this order was counterbalanced to ensure equal participants in both categories, VR first and FS first. A verbal instructions protocol was used to ensure that instructions were held constant for all participants. Before the start of the session, participants were informed about the study itself but not about the purpose or the hypotheses. Participants were asked to sign the consent form and fill out the ``participant demographics and pre-exposure'' questionnaire. This was followed by a brief introduction of their specific game and the objective they had to complete. Then, the Empatica E4 wristband was attached to the participants and switched on. The LED signals of the E4 device were noted to ensure correct functioning. After the equipment test was complete, the participants were told to relax and not think of anything extreme or arousing to establish a baseline. This lasted for 5 minutes. In the case of the VR experiment,  participants were introduced to the use of VR while the researchers helped fit the Rift S headset to the participant's comfort. Comfort in the use of hand controllers was confirmed by asking participants to navigate basic menus. In the case of the FS, the participants were shown the basic controls of the keyboard and mouse. In Dirt Rally 2.0, the Thrustmaster wheel was used in both FS and VR. 
 
 After this, the study continued with the participant playing the game, trying to complete the decided tasks. These tasks (and the game level/map) were the same for every participant playing the same game to ensure similar gameplay. The session ended when the participant completed the tasks for that specific game. At the end of each game session, participants filled out the ``Post-Exposure'' questionnaire consisting of the SAM and VAS. At the end of both sessions, participants were briefed about the study objectives and thanked for their participation.

\subsection{Dataset}
We use the FLIRT module\cite{flirt2021} in Python to extract various statistical and signal features for the EDA and HRV data. FLIRT module preprocessed the data for artifact removal and noise filtering using the extended Kalman filter (EKF) for EDA data and the Malik\cite{10.1093/oxfordjournals.eurheartj.a014868}, Kamarth\cite{kamath1995correction} and Acar\cite{ACAR2000123} rule for the heart rate. Accelerometer data is ignored in our analysis because it does not carry significant impressions useful for the final analysis. The trial was conducted on 36 participants; however, three of the data points had to be dropped due to participants' inability to complete the gameplay for various reasons. This led to a total of 33 participants, split between 4 games(see Table \ref{tab:dataset}).
We provide anonymized ACC, BVP, EDA, HR, IBI, and TEMP data in the form of CSV files for both VR and FS. Each CSV file starts with a timestamp, followed by the frequency of the recording and the data collected over the recording. We have also added the script we used to get the results of this study.

\begin{table}[!h]
\begin{center}
\caption{ Number of Participants and Average Time Taken for Different Games in the Dataset }
\label{tab:dataset}
\begin{tabular}
{m{0.01\textwidth} m{0.05\textwidth}m{0.05\textwidth} m{0.05\textwidth}m{0.06\textwidth} m{0.12\textwidth}}

     &  \centering Dirt Rally 2 & Phasmo-phobia & Minecraft & \centering War Thunder & Duration\\
    & \centering(HVHA) & \centering(LVHA) & \centering(LVLA) & \centering(HVLA) & (minutes)\\ \midrule
  VR & \centering 9 & \centering 8 & \centering 7 & \centering 9 & $\mu$= 13.7, $\sigma$ = 2.76\\
  FS & \centering 9 & \centering 8 & \centering 7 & \centering 9 & $\mu$= 14.1, $\sigma$ = 2.65\\
  \midrule
 \end{tabular}
\end{center}
\end{table}

\vspace{-15pt}
\section{Results}\label{Results}

\subsection{Validation of Stimulation}

Through our analysis, we intended to confirm that the chosen games indeed elicit the emotions assigned to them by the CMA quadrant. Table~\ref{tab:CMA_AV} provides the mean ($\mu$) and standard deviation ($\sigma$) for different games during FS gameplay, VR gameplay, as well as both types of gameplays taken together (VR+FS). We observe that the arousal and valence fall in the expected quadrant. However, further statistical significance analysis is required to confirm this. Considering both modalities taken together (VR+FS), we observe that higher reported arousal is evident for games rated as high arousal. (HVHA Median: 4.0 IQR:(4.0 - 5.0); LVHA Median: 3.5 IQR: (3.0 - 5.0). Similarly, low arousal is reported for games rated as Low arousal (HVLA Median: 3.0 IQR: (2.0 - 3.8)), LVLA Median: 2.0 IQR: (2.0 - 2.8)). This trend continues for valence as well with High valence rated games, generating a higher reported valence (HVHA Median:  4.0 IQR: (3.0 - 4.8); HVLA Median: 4.0 IQR: (3.0 - 4.0)) and vice-versa (LVHA Median: 3.0 IQR: (2.0 - 4.0); LVLA Median: 3.0 IQR: (2.2 - 4.0)). Taken independently, both the VR and FS modalities confer the above trends.\\

\begin{figure*}[!h]
    \centering
    \begin{tabular}{ccc}
    \begin{tikzpicture}
        \begin{axis}[ylabel={Arousal}, xlabel={CMA Quadrant}, xtick={0.5,1.5,2.5,3.5},
          xticklabels={LVHA, HVLA, LVLA, HVHA},
          every axis plot/.append style={fill,fill opacity=0.25},
          area legend, legend style={legend columns=-1, fill=gray!30!white, draw=none, at={(1,0)}, anchor=south east},
          width=0.32\textwidth,
          height=0.27\textwidth,
          boxplot={
            draw direction=y,
            box extend=0.2,
            every median/.style=ultra thick,
            draw position={
                    1/4 + floor(\plotnumofactualtype/2)
                      + 1/4*fpumod(\plotnumofactualtype,2)
                }
            },          
          x tick label style={
              text width=2.5cm,
              align=center
          },  
          cycle list={{brown},{black}}
          ]
            \addplot+ [boxplot prepared={
            lower whisker= 2,
		lower quartile=  2.75 ,
		median= 3,
		upper quartile=  3.25 ,
		upper whisker=  5
            },] coordinates{};
            \addplot+ [boxplot prepared={
            lower whisker= 3,
		lower quartile=  3.75 ,
		median= 5,
		upper quartile= 5,
		upper whisker=  5
            },] coordinates{};
            
            \addplot+ [boxplot prepared={
            lower whisker= 1,
		lower quartile= 2,
		median= 2,
		upper quartile=  2.5 ,
		upper whisker=  3
            },] coordinates{};
            \addplot+ [boxplot prepared={
            lower whisker= 3,
		lower quartile= 3,
		median= 4,
		upper quartile=  4.5 ,
		upper whisker=  5
            },] coordinates{};
            
            \addplot+ [boxplot prepared={
            lower whisker= 2,
		lower quartile= 2,
		median= 2,
		upper quartile= 3,
		upper whisker=  3
            },] coordinates{};
            \addplot+ [boxplot prepared={
            lower whisker= 1,
		lower quartile= 1,
		median= 2,
		upper quartile= 2,
		upper whisker=  4
            },] coordinates{};
            
            \addplot+ [boxplot prepared={
            lower whisker= 4,
		lower quartile= 4,
		median= 4,
		upper quartile= 4,
		upper whisker=  5
            },] coordinates{};
            \addplot+ [boxplot prepared={
            lower whisker= 3,
		lower quartile= 4,
		median= 5,
		upper quartile= 5,
		upper whisker=  5
            },] coordinates{};
            
        \end{axis}
    \end{tikzpicture} 
    &
    \begin{tikzpicture}
        \begin{axis}[ylabel={Valence}, xlabel={CMA Quadrant}, xtick={0.5,1.5,2.5,3.5},
          xticklabels={LVHA, HVLA, LVLA, HVHA},
          every axis plot/.append style={fill,fill opacity=0.25},
          area legend, legend style={legend columns=-1, fill=gray!30!white, draw=none, at={(1,0)}, anchor=south east},
          width=0.32\textwidth,
          height=0.27\textwidth,
          boxplot={
            draw direction=y,
            box extend=0.2,
            every median/.style=ultra thick,
            draw position={
                    1/4 + floor(\plotnumofactualtype/2)
                      + 1/4*fpumod(\plotnumofactualtype,2)
                }
            },          
          x tick label style={
              text width=2.5cm,
              align=center
          },  
          cycle list={{brown},{black}}
          ]
            \addplot+ [boxplot prepared={
            lower whisker= 2,
		lower quartile= 2,
		median= 3,
		upper quartile=  3.25 ,
		upper whisker=  4
            },] coordinates{};
            \addplot+ [boxplot prepared={
            lower whisker= 2,
		lower quartile=  2.75 ,
		median=  3.5 ,
		upper quartile= 4,
		upper whisker=  5
            },] coordinates{};
            
            \addplot+ [boxplot prepared={
            lower whisker= 2,
		lower quartile= 3,
		median= 4,
		upper quartile= 4,
		upper whisker=  5
            },] coordinates{};
            \addplot+ [boxplot prepared={
            lower whisker= 3,
		lower quartile=  3.5 ,
		median= 4,
		upper quartile=  4.5 ,
		upper whisker=  5
            },] coordinates{};
            
            \addplot+ [boxplot prepared={
            lower whisker= 2,
		lower quartile= 3,
		median= 3,
		upper quartile= 4,
		upper whisker=  5
            },] coordinates{};
            \addplot+ [boxplot prepared={
            lower whisker= 1,
		lower quartile= 2,
		median= 3,
		upper quartile= 4,
		upper whisker=  5
            },] coordinates{};
            
            \addplot+ [boxplot prepared={
            lower whisker= 2,
		lower quartile= 3,
		median= 3,
		upper quartile= 4,
		upper whisker=  4
            },] coordinates{};
            \addplot+ [boxplot prepared={
            lower whisker= 1,
		lower quartile= 4,
		median= 5,
		upper quartile= 5,
		upper whisker=  5
            },] coordinates{};
            
        \end{axis}
    \end{tikzpicture} 
    &
    \begin{tikzpicture}
        \begin{axis}[ylabel={Dominance}, xlabel={CMA Quadrant}, xtick={0.5,1.5,2.5,3.5},
          xticklabels={LVHA, HVLA, LVLA, HVHA},
          every axis plot/.append style={fill,fill opacity=0.25},
          area legend, legend style={legend columns=-1, fill=gray!30!white, draw=none, at={(1,0)}, anchor=south east},
          width=0.32\textwidth,
          height=0.27\textwidth,
          boxplot={
            draw direction=y,
            box extend=0.2,
            every median/.style=ultra thick,
            draw position={
                    1/4 + floor(\plotnumofactualtype/2)
                      + 1/4*fpumod(\plotnumofactualtype,2)
                }
            },          
          x tick label style={
              text width=2.5cm,
              align=center
          },  
          cycle list={{brown},{black}}
          ]
            \addplot+ [boxplot prepared={
            lower whisker= 2,
		lower quartile= 3,
		median=  3.5 ,
		upper quartile= 4,
		upper whisker=  5
            },] coordinates{};
            \addplot+ [boxplot prepared={
            lower whisker= 1,
		lower quartile= 2,
		median= 2,
		upper quartile= 3,
		upper whisker=  3
            },] coordinates{};
            
            \addplot+ [boxplot prepared={
            lower whisker= 2,
		lower quartile= 2,
		median= 2,
		upper quartile= 4,
		upper whisker=  4
            },] coordinates{};
            \addplot+ [boxplot prepared={
            lower whisker= 1,
		lower quartile= 1,
		median= 1,
		upper quartile= 2,
		upper whisker=  3
            },] coordinates{};
            
            \addplot+ [boxplot prepared={
            lower whisker= 1,
		lower quartile= 2,
		median= 3,
		upper quartile= 3,
		upper whisker=  4
            },] coordinates{};
            \addplot+ [boxplot prepared={
            lower whisker= 1,
		lower quartile= 2,
		median= 3,
		upper quartile= 4,
		upper whisker=  5
            },] coordinates{};
            
            \addplot+ [boxplot prepared={
            lower whisker= 1,
		lower quartile= 2,
		median= 2,
		upper quartile= 3,
		upper whisker=  4
            },] coordinates{};
            \addplot+ [boxplot prepared={
            lower whisker= 1,
		lower quartile= 1,
		median= 2,
		upper quartile= 3,
		upper whisker=  4
            },] coordinates{};
            
        \end{axis}
    \end{tikzpicture} 
    \\(a) & (b) &(c)\\
    &\includegraphics[width=0.15\textwidth]{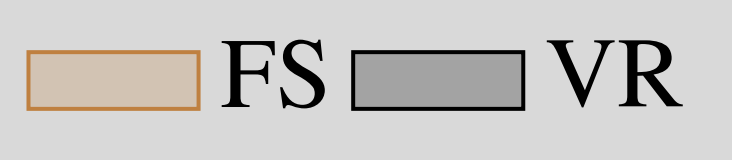}\\
    \end{tabular}
        \caption{Reported values of (a) arousal, (b) valence, and (c) dominance of players using SAM in different games in both VR and FS.}
        \label{fig:SAM_Comp}
\end{figure*}
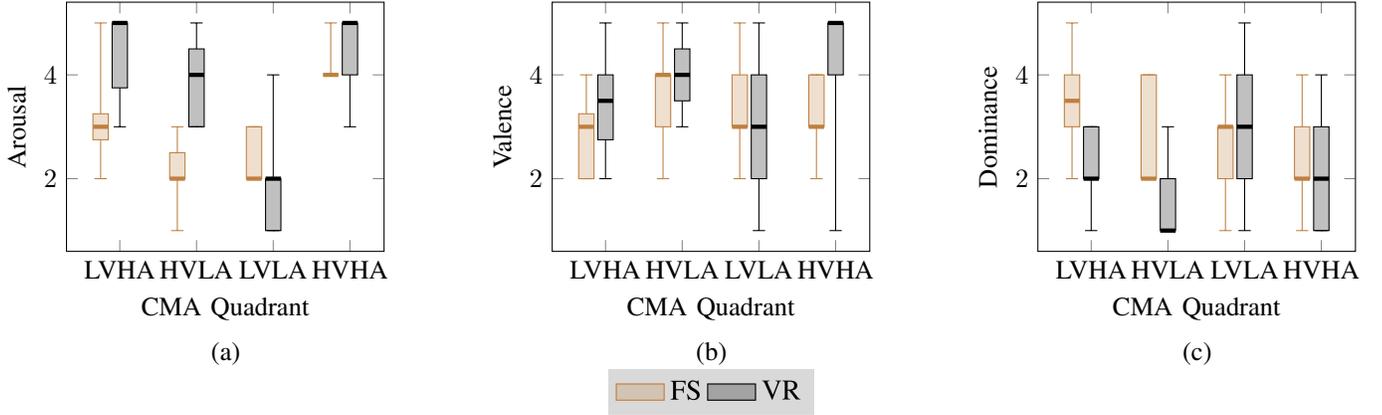

\begin{table}[h]

\caption{Reported Arousal and Valence Ratings According to the CMA Quadrant.}
\label{tab:CMA_AV}
\begin{center}
\begin{tabular}{p{0.1\textwidth} p{0.05\textwidth}p{0.025\textwidth} p{0.025\textwidth}p{0.025\textwidth} p{0.025\textwidth}p{0.025\textwidth} p{0.025\textwidth}} \toprule
    \textbf{Intended CMA} & \textbf{Type} & \multicolumn{2}{c}{\textbf{Arousal}} & \multicolumn{2}{c}{\textbf{Valence}} & \multicolumn{2}{c}{\textbf{Distance r}} \\
     \textbf{Quadrant}& & $\mu$ & $\sigma$ & $\mu$ & $\sigma$ & $\mu$ & $\sigma$\\
     \midrule
    & VR & 2.00 & 0.94 & 3.00 & 1.25 & 3.72 & 1.25 \\
    LVLA& FS & 2.33 & 0.47 & 3.56 & 0.96 & 4.27 & 0.98 \\
     & VR+FS & 2.17 & 0.76 & 3.28 & 1.15 & 4.00 & 1.16 \\
    \midrule
     & VR & 4.38 & 0.86 & 3.38 & 0.99 & 5.61 & 0.87 \\
    LVHA & FS & 3.12 & 0.93 & 2.88 & 0.78 & 4.36 & 0.70 \\
     & VR+FS & 3.75 & 1.09 & 3.12 & 0.93 & 4.99 & 1.01 \\
    \midrule
     & VR & 3.86 & 0.83 & 4.00 & 0.76 & 5.62 & 0.72 \\
    HVLA & FS & 2.14 & 0.64 & 3.57 & 0.90 & 4.24 & 0.79 \\
     & VR+FS & 3.00 & 1.13 & 3.79 & 0.86 & 4.93 & 1.02 \\
    \midrule
     & VR & 4.44 & 0.68 & 4.22 & 1.23 & 6.25 & 0.69 \\
    HVHA & FS & 4.22 & 0.42 & 3.22 & 0.79 & 5.36 & 0.53 \\
     & VR+FS & 4.33 & 0.58 & 3.72 & 1.15 & 5.81 & 0.76 \\
    \bottomrule
\end{tabular}
\end{center}
\end{table}

To ascertain the effect of the games on affective valence and arousal, a Kruskal-Wallis ANOVA was conducted on the data (Table \ref{tab:CMA_AV}) as it did not conform to the assumptions of parametric ANOVA. In the case of both FS and VR, the reported arousal varied significantly (Kruskal-Wallis H(3): 30.67, p-value: 9.97e-7) across the games in all four CMA quadrants.  The Kruskal-Wallis test result was followed up with posthoc Dunn's test the results are presented in the table. This indicates that the participants playing the games in the  LVHA  and  HVHA categories experience significantly higher arousal as compared to those playing games in the  LVLA  and  HVLA categories. Even though participants perceived the  HVHA  in FS as less arousing than in VR, the arousal ratings in this quadrant were still significantly higher than the  LVLA  of the same. In summary, the participants experienced four distinct emotional states over the differing valence and arousal dimensions (HVHA, HVLA, LVHA, LVLA) through a statistically significant difference in reported arousal and valence.

\subsection{Psychological Results}
\subsubsection{SAM}
We show plots of the mean and standard deviation of the reported arousal, valence, and dominance of different games in Fig.~\ref{fig:SAM_Comp}.  The data was found to be non-normal upon using the Shapiro-Wilks test (for all: W$>$0.89, p-value$<$0.00009). We conducted a nonparametric Wilcoxon's signed-rank test (W) for the analysis of nonparametric data, specifically used to evaluate matched-pair data based on their differences \cite{woolson2007wilcoxon}. The 'r' statistic serves as an effect size measure for this test, spanning a range from 0 to 1, wherein a value of 0 implies a minor effect and 1 denotes a substantial effect. Upon comparing the arousal between VR and FS gameplay using nonparametric Wilcoxon's signed-rank test(W), we found that the reported arousal in VR (Median: 4.00 IQR: (3.00 - 5.00)) was higher than FS (Median: 3.00 IQR: (2.00 - 4.00)), which was statistically significant according to the above test(W(33) = 66.5, p-value = 0.007, r = 0.47). We also observed a higher value for valence in VR gameplay (Median: 4.00 IQR: (3.00 - 5.00))  as compared to FS gameplay (Median: 3.00 IQR: (3.00 - 4.00))  as shown in Fig.~\ref{fig:SAM_Comp}(b). However, the Wilcoxon signed-rank test found this difference statistically non-significant (W(33) = 75.5, p-value = 0.262, r = 0.19). In contrast, we observed that dominance was lower in VR gameplay (Median: 2.00 IQR: (1.00 - 3.00))  compared to FS gameplay (Median: 3.00 IQR: (2.00 - 4.00)), and this difference was found to be statistically significant  (W(33) = 88.0, p-value = 0.02, r = 0.41). Finally, we compared the distance from the center of reported arousal and valence between VR and FS. We observed that this was much higher in VR (Median: 5.66 IQR: (5.00 - 6.40)) as compared to FS (Median: 4.47 IQR: (4.12 - 5.39)), this difference was found to be statistically significant upon conducting the Wilcoxon's signed-rank test (W(33) = 92.5, p-value=0.0039, r=0.044).

To ascertain the relationship between the reported arousal, valence, and dominance, a non-parametric Spearman's rank correlation was done, where we found that the dominance was weakly correlated to both arousal($\rho$(64) = -0.27, p = 0.029) and valence($\rho$(64) = -0.27, p = 0.031). Upon further inspection, we found a moderate negative association between dominance and valence for high-rated dominance cases (dominance $\geq$ 4, $\rho$(14) = -0.42, p = 0.12), which was not significant; however, there was also a strong negative correlation between the same for low-rated dominance cases (dominance $\leq$ 2, $\rho$(32) = -0.53, p = 0.01), which was statistically significant.

\subsubsection{VAS}
\begin{figure}[!h]
  \centering
  \begin{tabular}{c}
    \begin{tikzpicture}
        \begin{axis}[ylabel={VAS Score},
            xlabel={Emotion}, xtick={0, 1, 2, 3, 4, 5, 6, 7, 8, 9},
        xticklabel style={rotate=60},
          xticklabels={Joy,
            Anger,
            Calmness,
            Sadness,
            Disgust,
            Relaxation,
            Happiness,
            Fear,
            Anxiousness,
            Dizziness},
          every axis plot/.append style={fill,fill opacity=0.25},
          area legend, legend style={legend columns=-1, fill=gray!30!white, draw=none},
          width=0.5\textwidth,
          height=0.27\textwidth,
          boxplot={
            draw direction=y,
            box extend=0.25,
            every median/.style=ultra thick,
            draw position={
            \plotnumofactualtype/2
            - 1/4*mod(\plotnumofactualtype,2)
                }
            },          
          x tick label style={
              text width=2cm,
              align=center
          },  
          cycle list={{brown},{black}}
          ]
            \addplot+ [boxplot prepared={
            lower whisker=  0,
		lower quartile=  28,
		median=  55,
		upper quartile=  70,
		upper whisker=  100
            },] coordinates{};
            \addplot+ [boxplot prepared={
            lower whisker=  14,
		lower quartile=  50,
		median=  70,
		upper quartile=  85,
		upper whisker=  100
            },] coordinates{};
            
            \addplot+ [boxplot prepared={
            lower whisker=  0,
		lower quartile= 0,
		median=  100,
		upper quartile=  26,
		upper whisker=  65
            },] coordinates{};
            \addplot+ [boxplot prepared={
            lower whisker= 0,
		lower quartile= 2,
		median=  18,
		upper quartile=  40,
		upper whisker=  97
            },] coordinates{};
            
            \addplot+ [boxplot prepared={
            lower whisker= 0,
		lower quartile= 0,
		median=  24,
		upper quartile=  58,
		upper whisker=  100
            },] coordinates{};
            \addplot+ [boxplot prepared={
            lower whisker= 0,
		lower quartile= 1,
		median=  11,
		upper quartile=  50,
		upper whisker=  100
            },] coordinates{};
            
            \addplot+ [boxplot prepared={
            lower whisker= 0,
		lower quartile= 0,
		median= 0,
		upper quartile=  19,
		upper whisker=  48
            },] coordinates{};
            \addplot+ [boxplot prepared={lower whisker= 0,
		lower quartile= 0,
		median= 0,
		upper quartile=  10,
		upper whisker=  25
            },] coordinates{};

            \addplot+ [boxplot prepared={
            lower whisker= 0,
		lower quartile= 0,
		median= 0,
		upper quartile=  10,
		upper whisker=  25
            },] coordinates{};
            \addplot+ [boxplot prepared={
            lower whisker= 0,
		lower quartile= 0,
		median= 0,
		upper quartile=  15,
		upper whisker=  37
            },] coordinates{};
            
            \addplot+ [boxplot prepared={
            lower whisker= 0,
		lower quartile= 0,
		median=  30,
		upper quartile=  50,
		upper whisker=  75
            },] coordinates{};
            \addplot+ [boxplot prepared={
            lower whisker= 0,
		lower quartile= 0,
		median=  10,
		upper quartile=  40,
		upper whisker=  68
            },] coordinates{};
            
            \addplot+ [boxplot prepared={
            lower whisker=  0,
		lower quartile=  23,
		median=  51,
		upper quartile=  70,
		upper whisker=  100
            },] coordinates{};
            
            \addplot+ [boxplot prepared={
            lower whisker= 0,
		lower quartile=  28,
		median=  61,
		upper quartile=  83,
		upper whisker=  100.0
            },] coordinates{};
            
            \addplot+ [boxplot prepared={
            lower whisker= 0,
		lower quartile=  6,
		median=  20,
		upper quartile=  44,
		upper whisker=  100
            },] coordinates{};
            \addplot+ [boxplot prepared={lower whisker= 0,
		lower quartile=  6,
		median=  29,
		upper quartile=  54,
		upper whisker=  100
            },] coordinates{};

            \addplot+ [boxplot prepared={
            lower whisker= 0,
		lower quartile=  14,
		median=  33,
		upper quartile=  54,
		upper whisker=  100
            },] coordinates{};
            \addplot+ [boxplot prepared={
            lower whisker= 0,
		lower quartile=  14,
		median=  33,
		upper quartile=  54,
		upper whisker=  100
            },] coordinates{};
            
            \addplot+ [boxplot prepared={
            lower whisker= 0,
		lower quartile= 0,
		median= 0,
		upper quartile= 0,
		upper whisker=  31
            },] coordinates{};
            \addplot+ [boxplot prepared={
            lower whisker= 0,
		lower quartile=  6,
		median=  34,
		upper quartile=  65,
		upper whisker=  100
            },] coordinates{};
            
        \end{axis}
    \end{tikzpicture}  
    \\\hspace{1cm}\includegraphics[width=0.15\textwidth]{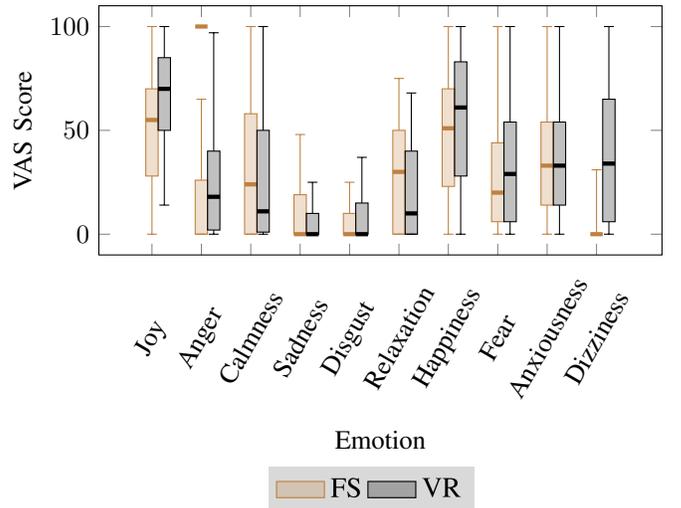}
  \end{tabular}
    \caption{Reported  values of different emotions upon playing games in VR and FS using VAS.}
    \label{fig:VAS_Comp}
\end{figure}

We plotted the mean and standard deviation recorded in VR and FS gameplay for the different emotions in Fig.~\ref{fig:VAS_Comp} and reported emotions per CMA quadrant. The data collected for the emotions were found to be non-normally distributed upon conducting the Shapiro-Wilks test (p-value $<$ 0.02). Upon conducting the Wilcoxon signed-rank test, we observe the p-values as shown in Table~\ref{tab:VAS_Pval}, with the significant values marked in bold. We observe that joy, anger, happiness, and dizziness are reported to be higher during VR gameplay as compared to FS gameplay. Of these, only joy and dizziness have a statistically significant difference between the two gameplay (joy: W(33) = 101; p-value = 0.004,  r = 0.04; dizziness: W(33) = 8.5, p-value = 0.0, r = 0.07) difference between the two gameplay. On the other hand, calmness, sadness, and relaxation, while higher during VR gameplay than in FS gameplay, are not different enough to be statistically significant.

\vspace{-10pt}

\begin{table}[h!]
\caption{A Comparison of Self Reported Levels of Different Emotions on the VAS Test between VR and FS Gameplay. Emotions with Statistically Significant Differences (p-value $<$ 0.05) Between VR and FS Gameplay are Shown in Bold.}
\label{tab:VAS_Pval}
\begin{center}
\begin{tabular}{c c c c c c c} \toprule
    \textbf{Emotion} & \multicolumn{2}{c}{\textbf{VR}} & \multicolumn{2}{c}{\textbf{FS}} & \textbf{$p$-value} & \textbf{r}\\
    & $\mu$ & $\sigma$ & $\mu$ & $\sigma$ &&\\ 
    \midrule
    \textbf{Joy} & \textbf{63.45} & \textbf{26.01} & \textbf{48.39} & \textbf{26.62} & \textbf{0.0041} & \textbf{0.04}\\
    Anger & 25.73 & 26.53 & 20.48 & 25.19 & 0.3198 & 0.02\\
    Calmness & 26.18 & 27.52 & 31.88 & 31.11 & 0.1389 & 0.02\\
    Sadness & 9.42 & 17.11 & 11.67 & 18.24 & 0.3007 & 0.02\\
    Disgust & 12.79 & 22.98 & 10.33 & 22.14 & 0.2176 & 0.02\\
    Relaxation & 23.30 & 26.88 & 28.24 & 27.16 & 0.1708 & 0.02\\
    Happiness & 53.55 & 33.28 & 46.64 & 30.38 & 0.1469 & 0.02\\
    Fear & 34.88 & 31.65 & 27.15 & 23.75 & 0.2096 & 0.0009\\
    Anxiousness & 37.55 & 31.58 & 35.27 & 26.79 & 0.9478 & 0.07\\
    \textbf{Dizziness} & \textbf{37.27} & \textbf{32.27} & \textbf{5.48} & \textbf{14.76} & \textbf{0.0000} & \textbf{0.53}\\
    \bottomrule
\end{tabular}
\end{center}
\end{table}

\begin{table}[h!]
    \centering
    \caption{Bonferroni-Corrected p-Values for Posthoc Dunn's Tests on SAM Data of all Four Quadrants}
    \label{tab:post-hoc}
    \begin{tabular}{ccccc}
        \toprule
        &  HVHA & HVLA & LVHA & LVLA\\
        \midrule
        HVHA &  1.0 & \textbf{0.002} & 0.156 & \textbf{1.3e-7} \\
        HVLA & \textbf{2.2e-3} & 1.0 & 0.1 & \textbf{0.006} \\
        LVHA & \textbf{0.015} & 0.1 & 1.0 & \textbf{2.1e-4} \\
        LVLA & \textbf{1.3e-7} & 0.06 & \textbf{2.2e-3} & 1.0\\
        \bottomrule
    \end{tabular}
\end{table}

\subsection{Physiological Results}
\subsubsection{EDA}
\begin{figure}[!h]
  \centering
  \begin{tabular}{c}
    \begin{tikzpicture}
        \begin{axis}[ylabel={SCL($\mu S$)}, xlabel={CMA Quadrant}, xtick={0.5,1.5,2.5,3.5},
          xticklabels={LVHA, HVLA, LVLA, HVHA},
          every axis plot/.append style={fill,fill opacity=0.25},
          area legend, legend style={legend columns=-1, fill=gray!30!white, draw=none},
          width=0.4\textwidth,
          height=0.27\textwidth,
          boxplot={
            draw direction=y,
            box extend=0.2,
            every median/.style=ultra thick,
            draw position={
                    1/4 + floor(\plotnumofactualtype/2)
                      + 1/4*fpumod(\plotnumofactualtype,2)
                }
            },          
          x tick label style={
              text width=2.5cm,
              align=center
          },  
          cycle list={{brown},{black}}
          ]
            \addplot+ [boxplot prepared={
            lower whisker=  0.33583577738689807 ,
		lower quartile=  1.9533892734883986 ,
		median=  3.33046988542287 ,
		upper quartile=  8.613471715076775 ,
		upper whisker=  19.068762973371193
            },] coordinates{};
            \addplot+ [boxplot prepared={
            lower whisker=  -1.2652777855027804 ,
		lower quartile=  0.966020249287771 ,
		median=  3.8146243263361326 ,
		upper quartile=  8.224880605947018 ,
		upper whisker=  19.07035674038908
            },] coordinates{};
            
            \addplot+ [boxplot prepared={
            lower whisker=  0.08732784758729119 ,
		lower quartile=  0.12616438918389966 ,
		median=  0.5302941801105288 ,
		upper quartile=  2.1521782768811946 ,
		upper whisker=  4.750626983354189
            },] coordinates{};
            \addplot+ [boxplot prepared={
            lower whisker=  0.06130901802814677 ,
		lower quartile=  0.15906866461084868 ,
		median=  0.28540184430673415 ,
		upper quartile=  0.9839411609232801 ,
		upper whisker=  10.379978162832021
            },] coordinates{};
            
            \addplot+ [boxplot prepared={
            lower whisker=  0.3077659401173776 ,
		lower quartile=  0.6518965795015969 ,
		median=  0.9811533797891319 ,
		upper quartile=  3.2174214459201633 ,
		upper whisker=  26.144398133418736
            },] coordinates{};
            \addplot+ [boxplot prepared={
            lower whisker=  0.6591670312115198 ,
		lower quartile=  1.5782878829169549 ,
		median=  2.0699324936377588 ,
		upper quartile=  3.2884425246918823 ,
		upper whisker=  4.104911503217235
            },] coordinates{};
            
            \addplot+ [boxplot prepared={
            lower whisker=  0.08051041821558683 ,
		lower quartile=  0.14597220898084035 ,
		median=  0.30927667018692334 ,
		upper quartile=  1.5827747943377697 ,
		upper whisker=  3.4403612237361294
            },] coordinates{};
            \addplot+ [boxplot prepared={lower whisker=0.000, 
            lower whisker=  -0.1494929625222622 ,
		lower quartile=  0.15128417063245905 ,
		median=  0.38740484903602934 ,
		upper quartile=  3.0193842134513504 ,
		upper whisker=  3.9194869950241746
            },] coordinates{};
            
        \end{axis}
    \end{tikzpicture} 
    \\
    \includegraphics[width=0.15\textwidth]{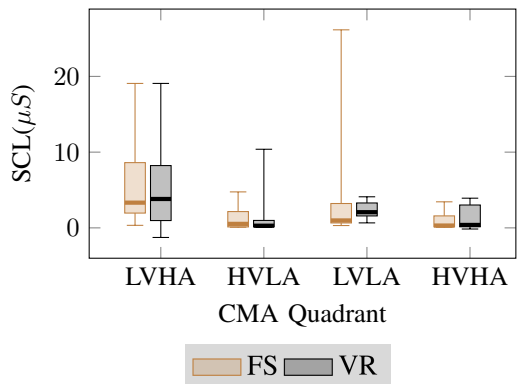}
  \end{tabular}
    \caption{ SCL of players in VR and FS to different games using EDA data.}
    \label{fig:SCL_By_Game}
\end{figure}

The raw EDA data is pre-processed using the convex EDA algorithm (cvxEDA) \cite{greco2015cvxeda} to obtain the tonic (SCL) and phasic (SCR) components.  The cvxEDA model is a physiologically inspired model that describes skin conductance as a combination of three factors: the phasic component, the tonic component, and additive white Gaussian noise. The cvxEDA algorithm uses convex optimization to break down the EDA signal into its different components and provides a comprehensive explanation of EDA.
 We utilized Z-score normalization to standardize the data around a mean of zero and a standard deviation of one. This method involved adjusting each feature by subtracting its mean and dividing by its standard deviation. By doing so, we achieved uniform scales across features, aiding interpretability, and analysis, by mitigating the impact of varying scales. From SCL, we obtain the statistical features such as mean, standard deviation, minimum, maximum, and percentiles shown in Fig.~\ref{fig:SCL_By_Game}. These results were found to be non-normal upon conducting the Shapiro-Wilks test (VR: p-value = 3.2e-12; PC: p-value=6.3e-9). While we observe higher SCL ratings during VR gameplay (Median: 1.53 IQR: (0.36-3.46)) than in FS gameplay (Median: 0.98 IQR: (0.31-3.22)), the difference between the two is statistically nonsignificant (W(33) = 275, p-value = 0.92, r = 0.001) according to the Wilcoxon Signed-rank test.
 
\begin{figure}[!h]
    \centering
    \begin{tabular}{c}
    \begin{tikzpicture}
        \begin{axis}[ylabel={Heart Rate($bpm$)}, xlabel={CMA Quadrant}, xtick={0.5,1.5,2.5,3.5},
          xticklabels={LVHA, HVLA, LVLA, HVHA},
          every axis plot/.append style={fill,fill opacity=0.25},
          area legend, legend style={legend columns=-1, fill=gray!30!white, draw=none}, legend pos= south west,
          width=0.4\textwidth,
          height=0.27\textwidth,
          boxplot={
            draw direction=y,
            box extend=0.2,
            every median/.style=ultra thick,
            draw position={
                    1/4 + floor(\plotnumofactualtype/2)
                      + 1/4*fpumod(\plotnumofactualtype,2)
                }
            },          
          x tick label style={
              text width=2.5cm,
              align=center
          },  
          cycle list={{brown},{black}},
          ]
            \addplot+ [boxplot prepared={
            lower whisker=69.130, 
		lower quartile=75.011, 
		median=79.866, 
		upper quartile=84.515, 
		upper whisker=89.330
            },] coordinates{};
            \addplot+ [boxplot prepared={
            lower whisker=76.394, 
		lower quartile=82.438, 
		median=89.809, 
		upper quartile=91.286, 
		upper whisker=95.353
            },] coordinates{};
            
            \addplot+ [boxplot prepared={
            lower whisker=69.810, 
		lower quartile=74.042, 
		median=75.818, 
		upper quartile=76.738, 
		upper whisker=81.975
            },] coordinates{};
            \addplot+ [boxplot prepared={
            lower whisker=64.031, 
		lower quartile=67.409, 
		median=73.199, 
		upper quartile=74.766, 
		upper whisker=75.521
            },] coordinates{};
            
            \addplot+ [boxplot prepared={
            lower whisker=57.411, 
		lower quartile=72.542, 
		median=78.369, 
		upper quartile=83.884, 
		upper whisker=87.565
            },] coordinates{};
            \addplot+ [boxplot prepared={
            lower whisker=68.166, 
		lower quartile=79.850, 
		median=84.123, 
		upper quartile=88.695, 
		upper whisker=98.180
            },] coordinates{};
            
            \addplot+ [boxplot prepared={
            lower whisker=59.671, 
		lower quartile=68.338, 
		median=76.459, 
		upper quartile=86.525, 
		upper whisker=104.856
            },] coordinates{};
            \addplot+ [boxplot prepared={lower whisker=61.180, 
		lower quartile=78.480, 
		median=84.740, 
		upper quartile=90.275, 
		upper whisker=101.442
            },] coordinates{};
            
        \end{axis}
    \end{tikzpicture} 
    \\ (a)\\
    \begin{tikzpicture}
        \begin{axis}[ylabel={LF/HF}, xlabel={CMA Quadrant}, xtick={0.5,1.5,2.5,3.5},
          xticklabels={LVHA, HVLA, LVLA, HVHA},
          every axis plot/.append style={fill,fill opacity=0.25},
          area legend, legend style={legend columns=-1, fill=gray!30!white, draw=none},
          width=0.4\textwidth,
          height=0.27\textwidth,
          boxplot={
            draw direction=y,
            box extend=0.2,
            every median/.style=ultra thick,
            draw position={
                    1/4 + floor(\plotnumofactualtype/2)
                      + 1/4*fpumod(\plotnumofactualtype,2)
                }
            },          
          x tick label style={
              text width=2.5cm,
              align=center
          },  
          cycle list={{brown},{black}},
          ]
            \addplot+ [boxplot prepared={
            lower whisker=0.620, 
		lower quartile=0.931, 
		median=1.061, 
		upper quartile=1.832, 
		upper whisker=3.136
            },] coordinates{};
            \addplot+ [boxplot prepared={
            lower whisker=0.908, 
		lower quartile=1.161, 
		median=1.688, 
		upper quartile=1.698, 
		upper whisker=2.105
            },] coordinates{};
            
            \addplot+ [boxplot prepared={
            lower whisker=0.833, 
		lower quartile=1.361, 
		median=1.638, 
		upper quartile=1.805, 
		upper whisker=1.998
            },] coordinates{};
            \addplot+ [boxplot prepared={
            lower whisker=0.729, 
		lower quartile=0.981, 
		median=1.258, 
		upper quartile=1.505, 
		upper whisker=1.896
            },] coordinates{};
            
            \addplot+ [boxplot prepared={
            lower whisker=0.446, 
		lower quartile=1.182, 
		median=1.459, 
		upper quartile=1.477, 
		upper whisker=1.941
            },] coordinates{};
            \addplot+ [boxplot prepared={
            lower whisker=0.316, 
		lower quartile=1.198, 
		median=1.907, 
		upper quartile=2.028, 
		upper whisker=2.888
            },] coordinates{};
            
            \addplot+ [boxplot prepared={
            lower whisker=0.477, 
		lower quartile=0.642, 
		median=0.807, 
		upper quartile=0.907, 
		upper whisker=1.008
            },] coordinates{};
            \addplot+ [boxplot prepared={lower whisker=0.664, 
		lower quartile=0.722, 
		median=0.802, 
		upper quartile=1.148, 
		upper whisker=2.006
            },] coordinates{};
            
        \end{axis}
    \end{tikzpicture} \\
    (b)\\
    \includegraphics[width=0.15\textwidth]{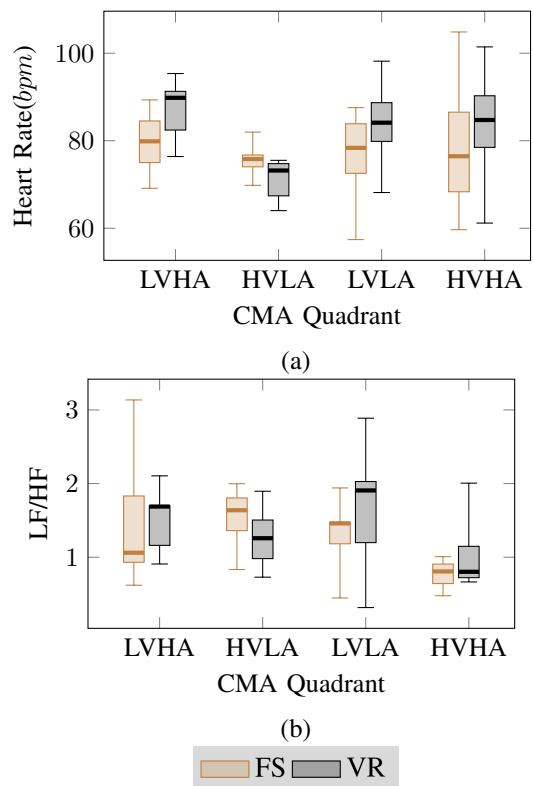}\\
     \end{tabular}

    \caption{ (a) Observed heart rates of participants playing different games in VR and FS. (b) Observed LF/HF features of participants playing different games in VR and FS.}
    \label{fig:HR_Comp}
\end{figure}

\subsubsection{Heart Rate}
We present the mean heart rate values for the participants during VR and FS gameplay, based on the game played in Fig.~\ref{fig:HR_Comp}(a).  The heart rate was normalized w.r.t. the baseline for each player using the Z-score normalization.  The heart rate observed during VR gameplay ($\mu$ = 81.52, $\sigma$ = 10.71) was higher than that observed during FS gameplay ($\mu$ = 77.27, $\sigma$ = 9.92). We found the difference to be statistically significant (t(32) = -2.5, p-value = 0.016, Cohen's d = 0.41) upon conducting a paired t-test on the data after confirming normality. We use Cohen's d as a measure of the effect size, 
it quantifies the difference between two groups or conditions in terms of their standard deviations, facilitating the comparison of effect sizes across different studies or experiments. \cite{cohen1992statistical}. Using the heart rate and the Inter Beat Interval (IBI), we extracted the LF and HF features. We then calculated the LF/HF ratio using these two and sorted them based on the different emotions as shown Fig.~\ref{fig:HR_Comp}(b).

 We found that the mean LF/HF values were also normally distributed according to the Shapiro-Wilks test (FS: p-value=0.21; VR: p-value=0.59). We also observed higher LF/HF values during VR gameplay ($\mu: $ 1.43, $\sigma$: 0.63) compared to FS gameplay ($\mu: $ 1.37, $\sigma$: 0.60). Upon conducting the Wilcoxon Signed-rank test, we observed that the difference was not statistically significant (t(24) = 0.33, p-value = 0.74, Cohen's d = 0.09).

\section{Discussion}\label{Discussion}

We present statistical validation of our six hypotheses presented before and provide inferences, followed by a discussion on the implications of our work and the limitations that we observed in our study.
\vspace{-5pt}
\subsection{Inferences}

We expected an elevated heart rate during VR gameplay according to our first hypothesis (H1) partly in response to the increased cognitive load in VR among other factors such as novelty. Our analysis observed a higher mean heart rate in the VR scenario.  A paired t-test indicates a  statistically significant result  (t(32) = -2.5, p-value = 0.016, Cohen's d = 0.41), validating H1.  Therefore, our results align with those of Malinska et al. \cite{malinska2015heart}, who explain this phenomenon as a consequence of increased immersion which may in turn lead to an additional cognitive burden on the user.

The second hypothesis (H2) proposes that increased immersion in VR gameplay will result in a greater elicitation of desired emotions. To evaluate this, we will refer to the radius of the SAM result in the CMA graph (Equation~\ref{eq:r}).  By examining this data, we can see that emotional elicitation (arousal) during VR gameplay (Median: 5.66 IQR: (5.00 - 6.40)) is higher than FS gameplay (Median: 4.47 IQR: (4.12 - 5.39)), and this difference is statistically significant (W(33) = 92.5, p-value = 0.003, r = 0.04) according to the Wilcoxon Signed Rank test. Furthermore, upon comparing the games within different quadrants of the CMA plot, we notice the discrepancy in distance from the center demonstrated statistical significance specifically in VR gameplay, exhibiting a substantially higher effect size (H(3): 10.7, p-value: 0.01, $\eta^2=0.35$) unlike the FS condition where statistical significance was not present (H(3): 6.3, p-value: 0.09, $\eta^2=0.21$)), using the Kruskal Wallis Test. 

According to our third hypothesis (H3), we expected that the LF/HF Sympathovagal balance would increase in the case of VR due to the increased emotional and physiological stress. In our study, we observed a higher value of LF/HF in VR  ($\mu: $ 1.43, $\sigma$: 0.63) as compared to FS ($\mu: $ 1.37, $\sigma$:0.60). Still, the difference was not statistically significant using a t-test (t(24) = -0.33, p-value = 0.74, Cohen's d = -0.09). This is most likely because the LF/HF ratio does not accurately measure cardiac sympathovagal balance \cite{billman_lfhf_2013}. 

The fourth hypothesis (H4) suggests that emotions felt by users will be less dominant in VR compared to FS displays.  Our experimental data backs this hypothesis; the dominance of emotions has been found to be lesser in VR gameplay (Median: 2.00 IQR: (1.00 - 3.00)) than in FS gameplay (Median: 3.00 IQR: (2.00 - 4.00)). This difference was significant, as indicated by the Wilcoxon signed-rank test (W(33) = 88.0, p-value = 0.02, r = 0.40). It is possible that this effect was due to the fact that many participants were new to VR and had little to no experience with it while they were familiar with using FS monitors. This lack of experience may have led to a decrease in emotional dominance in the unfamiliar VR environment.

According to our fifth hypothesis (H5), the participants should experience a higher level of arousal while in VR compared to being in a FS environment. Our study observed an increased SCL (VR Median: 1.17 IQR: (0.40-3.51); FS Median: 0.99 IQR: (0.31-3.28)); the difference was not statistically significant. However, we did observe a statistically significant increase in the participants' arousal levels when playing in VR (Median: 4.00 IQR: (3.00-4.00)) compared to FS (Median: 3.00 IQR: (2.00-4.00)) (W(33) = 66.5, p-value = 0.007, r = 0.04). This effect may be partially attributed to the greater immersion, intensity, variability, and dynamic nature of VR experiences.  We observe similar results with 3D VR eliciting a larger emotional response as compared to 2D VR\cite{estupinan2014can}.

 In line with our sixth hypothesis (H6), we found evidence of a correlation between dominance and arousal as well as dominance and valence. This supports previous research that reported a quadratic relationship between arousal and dominance \cite{warriner2013norms}. In addition, our results revealed a higher correlation between valence and dominance in VR ($\rho$(33) = 0.34, p = 0.047) compared to FS games ($\rho$(33) = 0.06, p = 0.73). We observe a similar trend between arousal and dominance (VR: $\rho$(33) = 0.31, p = 0.07; FS: $\rho$(33) = 0.002, p = 0.98). However, there is no correlation between arousal and valence, which might be due to the variance in the gameplay itself, and more research is needed about this effect.

\subsection{Implications and Future Work}
Our research provides valuable insights into user engagement and emotional response in both VR and FS gaming. These insights can be leveraged by developers to design platform-specific experiences that elicit intended emotions in end-users. Our findings can also inform the development of more engaging games for both entertainment and non-entertainment purposes, such as rehabilitation\cite{wiederhold2019physiological} and managing post-traumatic stress disorder\cite{reger2011effectiveness}.

By incorporating multimodal signals, developers can create adaptive gaming experiences that respond to users’ emotions and physiological responses, for example, an adaptive physical therapy game that adjusts to the user’s mental and physical state in real-time has immense potential \cite{yannakakis2018enhancing, Chanel2011}.

Our dataset can also be used by developers to design more engaging and immersive video games. Our findings indicate that emotional response is less dominant in VR compared to FS displays, suggesting that developers can explore ways to improve emotional engagement in VR games. By incorporating our findings into their game design, developers can enhance the emotional experience of players and improve the overall user experience.

While the physiological effects of VR have been investigated\cite{lavoie2021virtual}, the emotional reactions are not well-known. Our curated dataset, VRFS, can be used to further understand the emotional effects of gaming in VR and compare them with FS gaming as well as to validate the effect of arousal/valence on physiological signals. This knowledge can inform the development of a regulatory framework for gamification technology and help monitor and mitigate the emotional consequences of gaming.

\subsection{Limitations} 
One limitation of our study was the size and sample of the participants. Accordingly, a larger study is needed to produce a more extensive dataset that is more generalizable to a broader population. Our research focused exclusively on adults' emotional reactions, but it is very likely that younger players, including children, will experience more extreme emotional reactions \cite{drolet2007age}. Therefore, research should explore how different demographics, based on age, gender, etc., are susceptible to comparative effects. We acknowledge the differences in resolution (FS 1080p, VR 1440p), FPS (FS 60FPS, VR 80FPS), audio setup, and distance from the monitor in our experiment that arose due to the nature of VR setup as factors that may have influenced participant immersion which may be explored further. Another limitation of this study is that participants were asked to keep their off-hand as still as possible while playing the game to prevent motion artifacts. This may have reduced comfort for some participants, and it is possible that more natural data could have been collected if participants had been allowed to move their off-hand freely. The imposed restriction on off-hand movement during data collection introduces a partially ambulatory context, which could potentially impact the interpretation of LF/HF ratio. Therefore, it is important to consider the results within this specific context, and further research may be necessary to validate the use of LF/HF ratio in partially ambulatory conditions. It would also be interesting to explore individual differences in psychological and physiological responses, for example, based on openness to experience \cite{mccrae1992introduction}, exposure beyond the novelty effect period \cite{elor2021gaming}, previous experience with VR, etc. It will also be essential to gather longitudinal data, as this will enable a more robust understanding of how the observations of the current study change over time.

\section{Conclusion}\label{Conclusion}
In this study, we introduced a physiological dataset and performed a statistically significant analysis to compare emotional responses in VR and FS gaming. Using a pilot study with six participants, we curated four games designed to effectively evoke varied emotional responses. We captured physiological measures like blood volume pulse and electrodermal activity, alongside self-reported emotions from 33 participants. Our findings confirm that the selected games elicited the intended emotions. Furthermore, there was a distinct, statistically significant differentiation in arousal and valence ratings, showcasing that participants traversed varied emotional landscapes, notably HVHA, HVLA, LVHA, and LVLA. In a direct comparison, VR settings amplified emotional responses, with participants displaying heightened arousal, increased cognitive load, elevated stress, and reduced dominance when juxtaposed against FS gaming experiences. This research provides pivotal insights into the broader realm of virtual reality, elucidating emotional responses across varied applications such as gaming, and paves the way for comprehensive studies of VR's psychological implications. In contribution to the academic landscape, we're releasing the curated VRFS dataset, encompassing over 15 hours of multimodal physiological data contrasting FS and VR gaming across varied game genres. We believe this dataset will serve as an invaluable resource for the academic community, fostering future research endeavors in this domain.

\vspace{-5pt}
\section*{Acknowledgments}
This study is partially supported by the Center For Design and New Media (a TCS Foundation Initiative supported by Tata Consultancy Services) and the Infosys Centre for Artificial Intelligence at IIIT Delhi. We are grateful to all the participants for their cooperation during the study.

\bibliographystyle{IEEEtran}
\bibliography{VRFS-Main}


\begin{IEEEbiography}
[{\includegraphics[width=1in,height=1.25in,clip,keepaspectratio]{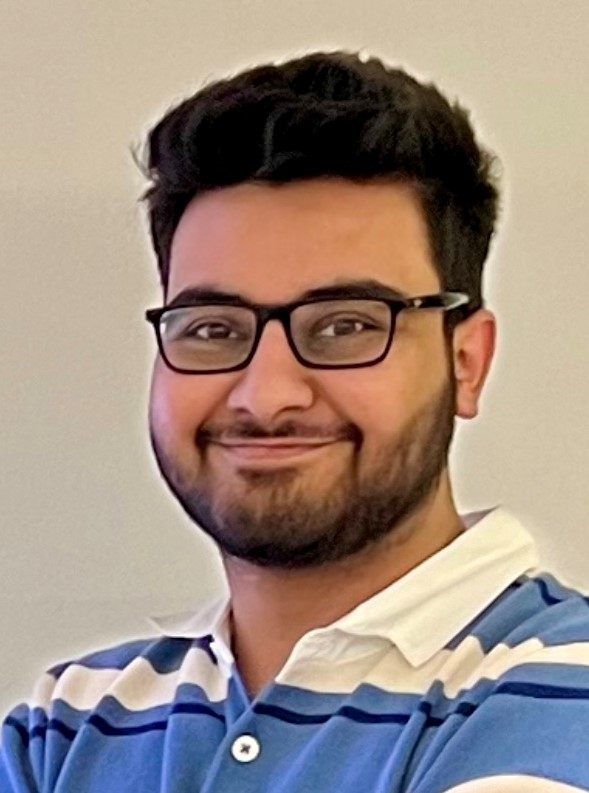}}]
{Ritik Vatsal}
Working as a Data Science Engineer at Airtel. Completed his B.Tech in Computer Science and Design with Honors from Indraprastha Institute of Information Technology, Delhi. He is the recipient of the Mitacs Globalink Research Scholarship 2022.  His areas of interest are Extended Reality, Machine Learning, Gaming, and Novel Interactions.
\end{IEEEbiography}

\vskip \baselineskip

\begin{IEEEbiography}
[{\includegraphics[width=1in,height=1.25in,clip,keepaspectratio]{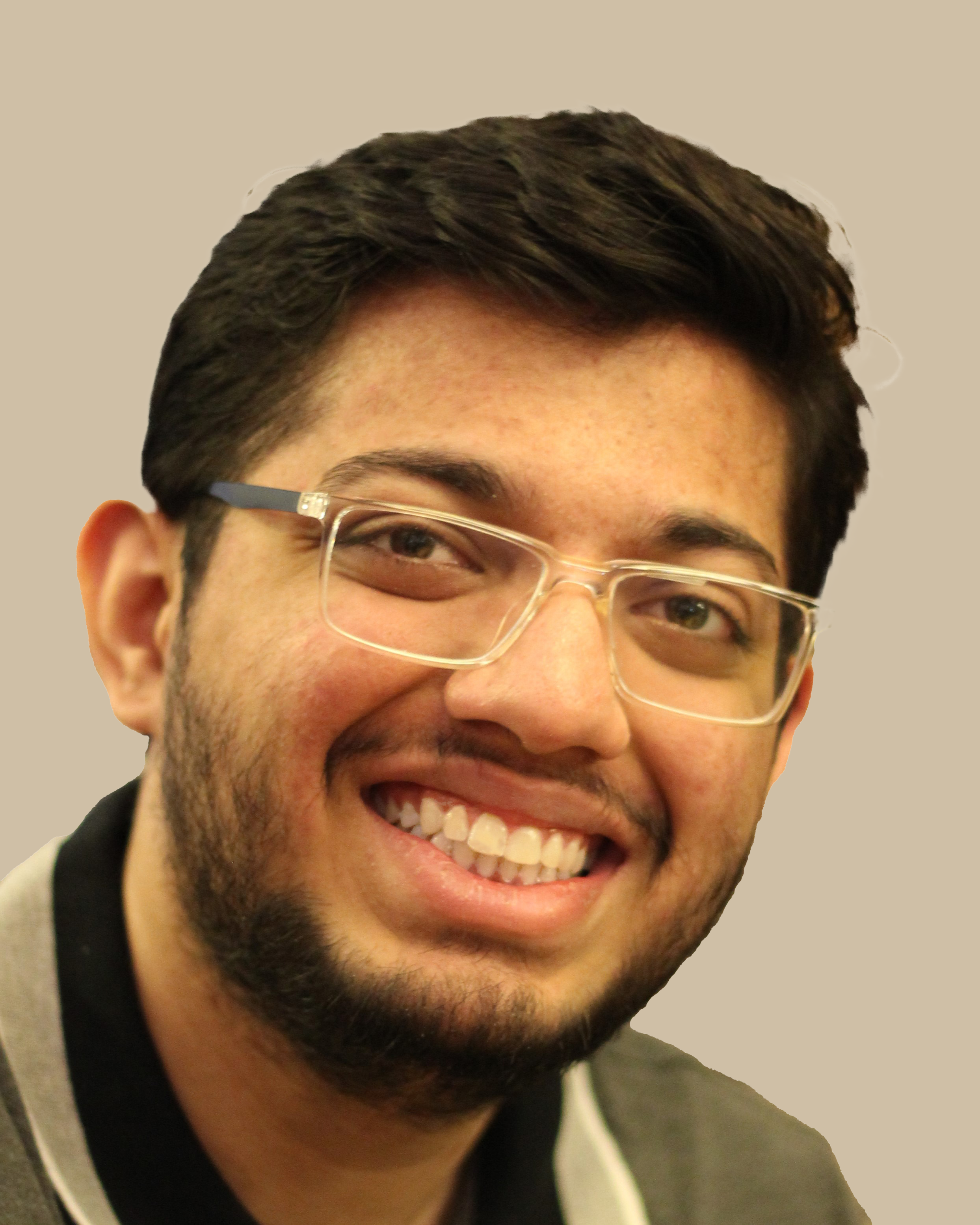}}]
{Shrivatsa Mishra}
 A graduate student at CU Boulder. Completed his B.Tech in Computer Science and Design from Indraprastha Institute of Information Technology, Delhi. He is the recipient of the Mitacs Globalink Research Scholarship 2022.  His areas of interest are HCI, Mixed Reality, Machine Learning, and Affective Computing.
\end{IEEEbiography}

\vskip \baselineskip

\begin{IEEEbiography}
[{\includegraphics[width=1in,height=1.25in,clip,keepaspectratio]{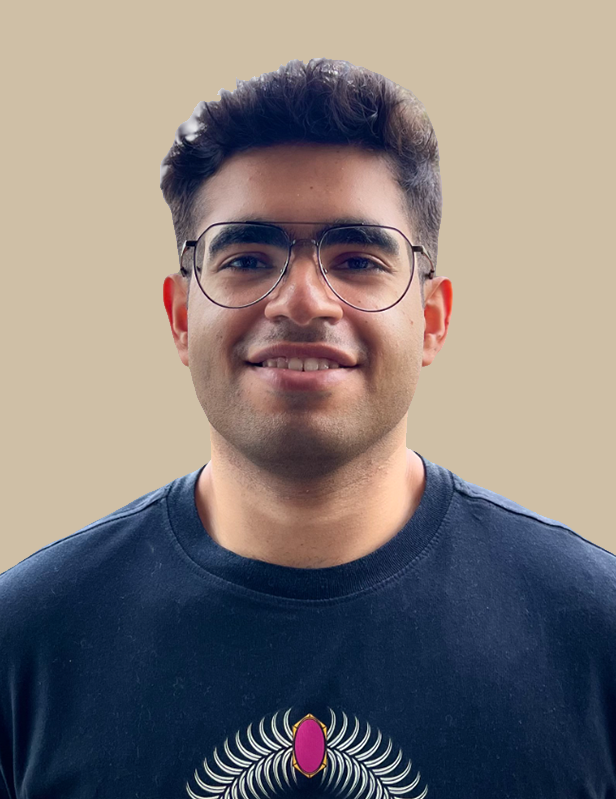}}]
{Rushil Thareja}
An Associate Data Scientist at Extramarks Education India. He holds a Bachelor's degree in Technology, specializing in Computer Science and Design, from IIIT Delhi. At the heart of his work is a deep-seated interest in multiple areas of research, including the application of AI in education, Generative AI, Language Models, and Multimodal Deep Learning. 
\end{IEEEbiography}

\vskip \baselineskip

\begin{IEEEbiography}
[{\includegraphics[width=1in,height=1.25in,clip,keepaspectratio]{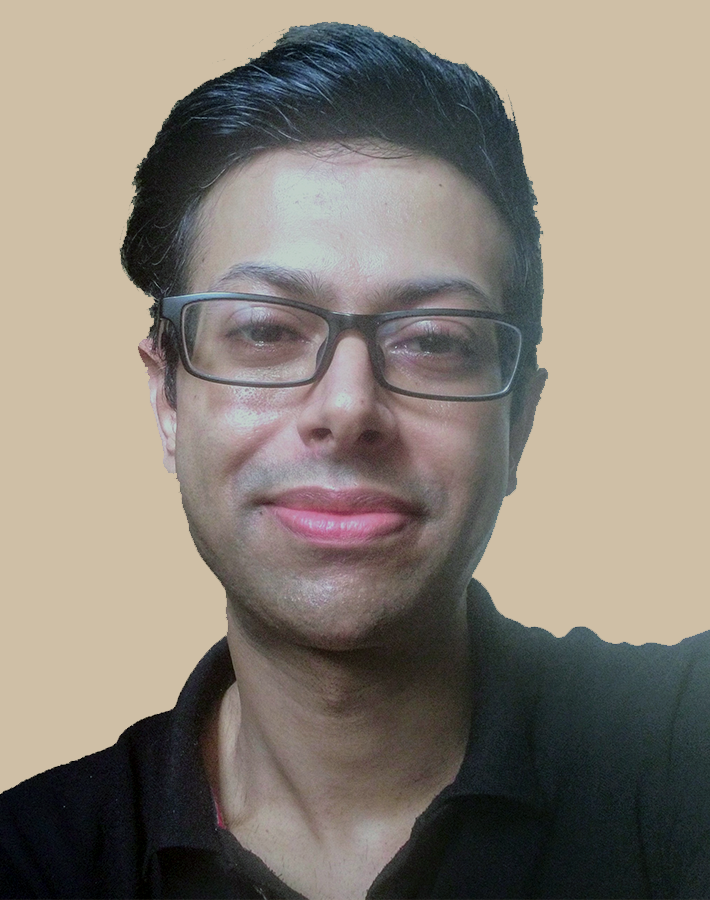}}]
{Mrinmoy Chakrabarty}
 Currently an Assistant Professor at Indraprastha Institute of Information Technology, Delhi. He completed his Ph.D. with Distinction from Osaka University Graduate School of Medicine, Japan in March 2017. His present experimental research employs psychophysics, psychophysiology and neuroimaging with computational data analysis to study cognitive functions in samples of clinical and healthy human populations.
\end{IEEEbiography}

\vskip \baselineskip

\begin{IEEEbiography}
[{\includegraphics[width=1in,height=1.25in,clip,keepaspectratio]{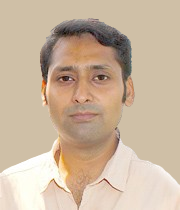}}]
{Ojaswa Sharma}
 Currently an Associate Professor at Indraprastha Institute of Information Technology, Delhi. He completed his Ph.D. in mathematics and computer science from the Technical University of Denmark, Denmark. His research spans various aspects of computer graphics, and computational geometry with focus on geometry creation and reconstruction, Virtual/Augmented Reality (AR/VR), volume rendering, and high-performance computing on GPU.
\end{IEEEbiography}

\vskip \baselineskip 

\begin{IEEEbiography}
[{\includegraphics[width=1in,height=1.25in,clip,keepaspectratio]{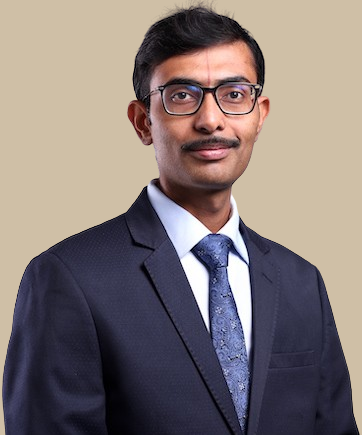}}]
{Jainendra Shukla}
 Ph.D. (URV, Spain), is a faculty at IIIT-Delhi and an Associate Editor at IEEE Transactions on Affective Computing. Motivated by Affective Computing, Social Robotics, and Ubiquitous Computing, he is enthusiastic about empowering machines with emotional intelligence and adaptive interaction skills, with the aim of enhancing the quality of life in healthcare and social settings.
\end{IEEEbiography}

\end{document}